\documentclass[onecolumn,prd,amsmath,amsssymb,groupedaddress,12pt]{revtex4}
\usepackage{graphicx}
\usepackage{dcolumn}
\usepackage{bm}
\usepackage{amssymb}
\def\lsim{\mathrel{\mathstrut\smash{\ooalign{\raise2.5pt\hbox{$<$}\cr\lower2.5pt\hbox{$\sim$}}}}}
\def\gsim{\mathrel{\mathstrut\smash{\ooalign{\raise2.5pt\hbox{$>$}\cr\lower2.5pt\hbox{$\sim$}}}}}
\def\be{\begin{equation}}
\def\ee{\end{equation}}
\def\bea{\begin{eqnarray}}
\def\eea{\end{eqnarray}}
\def\lf{\left (}
\def\rt{\right )}

\def\cO{{\cal O}}
\def\cI{{\cal I}}
\def\cQ{{\cal Q}}

\def\cD{{\cal D}}
\def\cC{{\cal C}}

\def\om{\omega}
\def\beq{\begin{eqnarray}}
\def\eeq{\end{eqnarray}}

\def\lsim{\mathrel{\rlap{\lower3pt\hbox{\hskip0pt$\sim$}}
    \raise1pt\hbox{$<$}}}         
\def\gsim{\mathrel{\rlap{\lower4pt\hbox{\hskip1pt$\sim$}}
    \raise1pt\hbox{$>$}}}         

\begin{document}

\title{Degravitation of the Cosmological Constant and Graviton Width}
\author{Gia Dvali$^{1}$, Stefan Hofmann$^{2}$ and Justin Khoury$^{2}$}

\affiliation{$^1$ Center for Cosmology and Particle Physics, Department of Physics,
New York University, New York, NY 10003\\
$^2$Perimeter Institute for Theoretical Physics, 
31 Caroline St. N., Waterloo, ON, N2L 2Y5, Canada}

\begin{abstract}
\begin{center}
{\bf Abstract}
\end{center}
\noindent  
We study the possibility of decoupling gravity from the vacuum energy. 
This is effectively equivalent to promoting Newton's constant to a
high-pass filter that degravitates 
sources of characteristic wavelength larger than a certain macroscopic
(super) horizon scale $L$.   We study the underlying physics and the consistency of this 
phenomenon.  In particular,  the absence of ghosts, already at
the linear level, implies that in any such theory the graviton should
either have a mass $1/L$, or be a resonance of similar width. 
This has profound physical implications for the degravitation idea.   

\end{abstract}

\maketitle

\section{Orientation}

Almost all of the effort in solving the cosmological constant problem~\cite{weinberg} has focused on the 
question --- {\sl why is the vacuum energy so small}?  However, since nobody has ever measured the energy 
of the vacuum by any means other than gravity, perhaps the right question to ask is --- 
{\sl why does the vacuum energy gravitate so little}~\cite{dgs, addg}?

 In Einstein's General Relativity (GR),  in which the messenger of the gravitational interaction at 
 observable macroscopic distances is a massless spin-2 particle, the universality of the graviton coupling automatically follows from gauge invariance.
Thus the two questions are equivalent.  
   
 However, the story is  {\it a priori} different in generally-covariant theories in which four-dimensional gravity is not mediated by a zero mode,  
 but by an effectively massive or resonance graviton~\cite{dgs}.  This is the story in theories with infinite-volume extra dimensions, such as the brane-world~\cite{DGP}
 (DGP) model. And in such theories the vacuum energy can indeed gravitate differently than other sources. Hence, the two questions can be distinguished. 
The same is true in any other generally covariant theory in which macroscopic gravity is not mediated by a massless graviton.   
 
Thus, in the context of large distance modified gravity theories one can ask whether the vacuum energy appears to be small because 
it is effectively {\sl degravitated}\,?  A phenomenological equation describing such a behavior in the four-dimensional language is~\cite{addg}
\begin{equation}
\label{central1}
 G_{\rm N}^{\; -1}(L^2\square) \mathcal{G}_{\mu\nu} \, = \,8\pi\,  T_{\mu\nu}, 
\end{equation}
where $\mathcal{G}_{\mu\nu}$ denotes the usual Einstein tensor, and $T_{\mu\nu}$ is the energy-momentum source. 
The only difference between (\ref{central1}) and the usual Einstein equations
is that Newton's constant $G_N$ is promoted to a differential operator $G_N(L^2\square)$. Thus Newton's constant is a function of the covariant 
d'Alembertian operator and acts as a high-pass filter. Here, $L$ denotes the filter scale, {\it i.e.}, the macroscopic distance scale that determines the passage.  
 Sources characterized by wavelengths~$\ll L$ pass undisturbed through the filter and gravitate normally. 
 Sources with characteristic wavelengths~$\gg L$, however, such as the cosmological constant, 
 are filtered out and effectively degravitated.
  
   In what follows, we shall discuss the consistency of the effective equation~(\ref{central1}) and 
  the fundamental physics behind it. 
   
   \subsection{Filtering and Graviton Mass/Width}
   
   One of the central points of the present paper will be that~(\ref{central1}) cannot represent a  
   consistent theory of massless gravitons with only two degrees of freedom. Instead it should 
   be understood as a well-defined limit of a theory in which the graviton is either a massive 
   spin-2 state with mass~$\sim 1/L$, or a resonance --- a sharply peaked superposition of an infinite number of 
   massive spin-2 states with tiny masses~$\sim 1/L$. 
 
    Indeed, looking at~(\ref{central1}) one quickly arrives at a puzzle, already present 
    at the linearized level:       
     \begin{equation}
\label{linear}
 \left ( 1 \, - \, {m^2(\square)\over \square} \right ) \mathcal{E}^{\;\; \alpha\beta}_{\mu\nu} \tilde{h}_{\alpha\beta}\, 
 = \, - T_{\mu\nu}\,,
\end{equation}
where
\begin{equation}
\label{einstein}
\mathcal{E}^{\; \; \alpha\beta}_{\mu\nu} h_{\alpha\beta}\, = \, \Box h_{\mu\nu} \, - 
\,\eta_{\mu\nu}  \Box h  \, - \, \partial_{\mu}\partial^{\alpha} h_{\alpha\nu} \, -\, 
\partial_{\nu} \partial^{\alpha} h_{\alpha\mu} \, + \, 
\eta_{\mu\nu} \partial^{\alpha}\partial^{\beta}h_{\alpha\beta}\, + \, \partial_{\mu}\partial_{\nu} h
\end{equation}
derives from the linearized Einstein tensor, and $h \equiv  \eta_{\alpha\beta}h^{\alpha\beta}$ as usual. 
Moreover, we have rewritten the filter into a more convenient  form 
\begin{equation}
\label{gnewton}
16\pi G_{\rm N}(L^2\square) \, = \, \left ( 1 -  {m^2(\square) \over \square} \right )^{-1}\,, 
\end{equation}
where $m^2(\square)$ is an appropriate function that encodes the filter scale $L$.  (Unitarity and causality 
constraints on $m^2(\square)$ will be discussed in detail below but are not important for the moment.)
At the linearized level, the Bianchi identity is automatically satisfied. Indeed,~(\ref{gnewton}) follows from the gauge-invariant action 
\begin{equation}
\label{linearaction1}
S
\,=\,
\int \, d^4 x \,~~ {1 \over 2} \, \tilde{h}^{\mu\nu}  
\left ( 1 \, - \, {m^2(\square)\over \square} \right ) \mathcal{E}^{\;\;\alpha\beta}_{\mu\nu} \tilde{h}_{\alpha\beta}\, + \, 
\int \, d^4 x \,~~ \tilde{h}^{\mu\nu} \,T_{\mu\nu}\,. 
\end{equation}

The puzzle is the following. Let us first take $m^2 \, =\, L^{-2} = {\rm constant}$, in which case
$\tilde{h}_{\mu\nu}$ describes a massive tensor particle. Nevertheless, it still propagates
two degrees of freedom. This is impossible. Indeed, the one particle exchange amplitude  mediated by $\tilde{h}$ between 
two conserved sources, $T_{\mu\nu}$ and $T_{\mu\nu}'$, is given by
\begin{equation}
\label{amplitude2}
({\rm Amplitude})_{\tilde{h}} \, \propto \, {T'_{\mu\nu}T^{\mu\nu} \, - {1 \over 2} T\; T' \,.
\over  \square \, - \, m^2 }\,,
\end{equation}
This continuously recovers the massless graviton result in the limit $m^2 \rightarrow 0$. However, it is well known that in Pauli-Fierz (PF) theory~\cite{FP}, the only consistent linear theory of massive gravity, 
the graviton carries three extra polarizations --- two helicity-1 and 
one helicity-0 states. The helicity-0 state couples to the trace of $T_{\mu\nu}$ and 
creates an extra scalar attraction, leading to the famous van Dam-Veltman-Zakharov  (vDVZ) discontinuity~\cite{vDVZ}  
at the linear level. This extra state would contribute the following missing term in the amplitude (\ref{amplitude2}), 
\begin{equation}
\label{amplitude0}
\Delta({\rm Amplitude}) \, \propto \, {1 \over 6} {  T\; T'
\over  \square \, - \, m^2}\,,
\end{equation}
which is the source of the vDVZ discontinuity in the $m^2 \, \rightarrow \, 0$ limit.   
So it appears that the action~(\ref{linearaction1}) cannot possibly describe a consistent theory of a  spin-2 particle!

Notice that promoting $m^2$ into a function $m^2(\Box)$ does not help, because in this case 
the amplitude can be spectrally expanded in a continuum of massive amplitudes
in the following way
\begin{equation}
\label{spectral1}
 \,{1 \over  \square \, - \, m^2 } = \, \int_0^{\infty} \, {\rm d}M^{2} \; { \rho (M^2,m^2) \over \square - M^2}\,,
\end{equation}
where $\rho(M^2,m^2)$ is the spectral function. Then, to each constituent amplitude the same objection 
applies as for the constant mass case. 

 The resolution of the above puzzle is simple. As it stands, the theory~(\ref{linearaction1}) is incomplete
 and must be supplemented by the missing amplitude (\ref{amplitude0}).  In other words, the theory 
 (\ref{linearaction1}) can only be understood as an effective theory describing 
 only the helicity-2 part  $\tilde{h}_{\mu\nu}$ (two polarizations) 
 of the massive (or resonance) graviton $h_{\mu\nu}$,  once the additional three extra degrees of freedom have been  
 integrated out.
 In Sec.~\ref{newdof} we shall prove this statement explicitly following the analysis of~\cite{gd}.
   
  Summarizing, any consistent theory that can degravitate the cosmological constant 
  must be equivalent at the linear level to a theory of massive (or resonance) gravity.   

\subsection{Goldstone-St\"uckelberg Story}

The inevitability of the graviton mass (or width) puts the degravitation of the cosmological constant in a completely 
 new perspective.  One may wonder whether this phenomenon can be understood in the language of 
 some `no-hair' argument.  Indeed, in a generally covariant theory of massive gravity
the metric fluctuation $h_{\mu\nu}$ itself must become observable, because of the underlying
 exact gauge invariance maintained by the extra polarizations.  These extra degrees of freedom 
 act like St\"uckelberg fields in the massless theory, {\it i.e.}, they promote the gauge field to an observable
 by rendering it gauge-invariant. 
 
The full metric $h_{\mu\nu}$ can be written in the St\"uckelberg  language as
 \begin{equation}
\label{stuckel}
h_{\mu\nu} \, = \, \hat{h}_{\mu\nu} \, + \, \partial_\mu A_\nu  + \partial_\nu A_\mu \, + \, ...\; ,
\end{equation}
where the St\"uckelberg vector $A_\mu$  guarantees the gauge invariance of $h_{\mu\nu}$ under the 
gauge transformation of $\hat{h}_{\mu\nu}$.  This is very similar to electrodynamics  in the Higgs (or Proca) phase, 
in which the photon $A_{\mu}$ becomes a gauge-invariant observable 
 \begin{equation}
\label{stuckelphoton}
A_{\mu} \, = \, \tilde{A}_{\mu} \, + \frac{1}{m}\partial_{\mu} \phi\,,
\end{equation}
where the appearance of the photon mass in the denominator follows from the canonical normalization 
of $\phi$. Indeed, $A_{\mu}$ is gauge-invariant since the gauge shift of its helicity-1 part, $\tilde{A}_{\mu} \,  \rightarrow \, \partial_{\mu} \omega$,
is exactly compensated by the corresponding shift in the Goldstone-St\"uckelberg field, $\phi \, \rightarrow \, \phi \, - \, m\omega$. 

Following this analogy, we are dealing with theories of gravity in the Higgs phase.  
The fact that in this phase $h_{\mu\nu}$ becomes a (gauge-invariant) physical observable is the key point in 
understanding the degravitation of the cosmological constant. Indeed, as any other physical observable,  
$h_{\mu\nu}$ cannot grow unbounded. Hence,   
any positive cosmological constant must eventually degravitate, provided the underlying theory is consistent. The latter issue will preoccupy much of our discussion. 

Therefore, degravitation of the cosmological constant in Higgsed gravity is analogous  to screening (or de-electrification) 
of an electric field in the Higgs vacuum in Maxwell theory, which takes place even in the presence 
of a uniform charge density. 

   As we will see,  the parallel between spin-1 and spin-2 theories goes surprisingly far. 
 Starting with massless electrodynamics, we demonstrate an analogue of the cosmological constant problem --- 
 a uniform charge density produces an electric field that grows unbounded.
 This problem is cured in the  Proca phase due to screening of the 
 background source.  By integrating out the  Goldstone-St\"uckelberg field, we find that screening is equivalent 
 to filtering. Furthermore, this filtering effect persists even in non-linear completions of Proca theory.
  
  Each step can be repeated for the spin-2 case.  
  First, we show that the cosmological constant  problem exists already at the linear level
  in the gauge-invariant theory of massless gravitons.  We then argue that the problem is cured by assigning a mass to the 
  gravitational field, 
  which effectively 
degravitates the source.  After integrating out the extra polarizations --- corresponding to Goldstone-St\"uckelberg fields for spin-2---, 
degravitation is shown to be equivalent to filtering, just as in the spin-1 case. 

 The conclusion is that, as long as we insist on exact abelian gauge shift symmetry, the analogy between spin-1 and spin-2 
 fields is perfect, both at the linear and non-linear levels. In both theories, uniform sources are screened or de-gravitated, and in both cases the effect can be understood as
 a filtering phenomenon,
 after the St\"uckelberg fields have been integrated out.  An important difference, however, is brought by 
 instabilities that are present in the massive spin-2 case, as discussed below. 
 
 \subsection{Role of Strong Coupling}
     
 Promoting the abelian gauge shift symmetry of the spin-2 field into the full non-abelian general covariance
 inevitably gives rise to a strong coupling regime for the extra polarizations
 in massive~\cite{ddgv, strong} or generic resonance~\cite{gd} theories of gravity.
 This phenomenon  also takes place in  DGP~\cite{ddgv,luty} , which from the effective field theory point of view can be considered as a special case of the resonant graviton with $m^2(\square) \, = \, \sqrt{\square} /L$  --- see, e.g.,~\cite{dgz}. For discussions on strong coupling in quantum theory, see~\cite{newgia,nicolis}.
  
    As we will see, a strong coupling regime may already be present in
  abelian spin-2 and spin-1 gauge theories with non-linear interactions, but in the non-abelian 
  spin-2 case it is inevitable and of universal nature.  
  
  The origin of the strong coupling regime can be understood in the following way. 
  At the linear level, the helicity-0 component $\chi$ contributes to $h_{\mu\nu}$ as
\begin{equation}
\label{opart}
h_{\mu\nu} \, = \, \frac{1}{3} {\partial_\mu \partial_\nu  \over m^2}\chi\, + \, ...\,,
\end{equation}
where the $1/m^2$ factor is due to the fact that the canonical kinetic term for $\chi$ stems from the graviton mass term.  
This means that  any nonlinear interaction of $h_{\mu\nu}$ that is not suppressed by an appropriate power of $m^2$  will 
generate an effective coupling for $\chi$ that is singular in the $m^2 \rightarrow  0$ limit. 
For example, a non-derivative quartic interaction of  $h_{\mu\nu}$ gives rise 
to a derivative quartic interaction of $\chi$:
\begin{equation}
\label{4int}
\lambda (h_{\mu\nu}h^{\mu\nu})^2 \, +\, ...\, \rightarrow \, 
{\lambda \over 81\;m^8} \left[(\partial_\mu \partial_\nu \chi)(\partial^{\mu}\partial^{\nu}\chi )\right]^2\, + ...
\end{equation}
Unless we assume $\lambda \propto m^8$, this term is singular in the $m^2 \rightarrow 0$ limit.
As a consequence, unlike other polarizations, $\chi$ becomes strongly coupled for momenta much below
the Planck scale.

The same holds true for the massive spin-1 vector field,  which can be rendered gauge-invariant
  in terms of the  {\sl canonically normalized}  Goldstone-St\"uckelberg field $\phi$ --- see~(\ref{stuckelphoton}). 
  The quartic spin-1 self-interactions then give rise to a strong coupling regime for $\phi$:
\begin{equation}
\label{4gold}
\lambda (A_{\mu}A^{\mu})^2 \, \rightarrow \, 
{\lambda \over m^4}\left[(\partial_\mu \phi)(\partial^{\mu}\phi )\right]^2\,.
\end{equation}
 Although the abelian case is similar for both spins, 
 the gauge-invariant completion with respect to full general covariance
 bears the following important difference.  In the abelian case, the coefficients in front of 
 non-linear  terms, such as $\lambda$ in (\ref{4int}) and (\ref{4gold}), can be freely chosen
 to vanish as appropriate powers of $m^2$. Hence, the strong  coupling may or may not be present in the theory.  
 In the case of non-abelian general covariance, however, there is no such choice 
 since the strong coupling phenomenon comes directly from the non-abelian completion of the mass term. 
 
  The strong coupling of longitudinal polarizations gives rise to the concept of
$r_\star$-radius, which plays a central role in the viability
of large distance modified gravity theories.  The scale $r_\star$ marks the length scale below which the extra polarizations become strongly coupled
due to their non-linear self-interactions. This is the case, for instance, near gravitating sources or in regions of high curvature. 
To be more precise, in regions where the curvature scale exceeds the filter scale $L^{-2}$, the 
contribution to the full metric from the extra polarizations into the metric is suppressed by powers of $L$.  
For instance, in cosmological backgrounds the concept of $r_\star$ tells
us that the metric is almost Einsteinian, with small corrections from the extra polarizations,
as long as the curvature of the Universe is much larger than $L^{-2}$.
These corrections only become important once the Hubble parameter $H$ drops to $\sim L^{-1}$. 
This universal behavior has been confirmed, both by general arguments~\cite{ddgv} and known exact cosmological~\cite{cedric} and static~\cite{andrei, lue, greg, nicolis,domainwalls}
solutions  in DGP.

In order to understand the role of the $r_\star$-phenomenon in degravitation, we study its effect in simple spin-1 and spin-2 theories
that exhibit filtering, in the sense that constant sources produce a pure gauge field with zero field strength. In particular, it is useful to analyze this relation in the 
decoupling limit, analogous to the one discussed in~\cite{strong}. In this limit, the gauge 
and St\"uckelberg fields no longer mix, and the latter form an independent non-linearly interacting system.  Taking such a limit, in massive spin-1 and spin-2 gauge theories we recover 
the $r_\star$-effect both for sources localized in space, as well for transitions localized in time.

   Recovery of the Einsteinian metric near spherically symmetric sources  in massive gravity 
   was first observed by Vainshtein~\cite{arkady}. However,  it was later understood that  in massive gravity
   such a recovery does not guarantee the
   existence of any sensible static solution interpolating between the linear and nonlinear regimes, {\it i.e.},  that matches to $1/r$ behavior at distances $r \gg r_\star$.  This issue is related to
   non-linear ghost instabilities and  goes back to 
 Boulware and Deser  (BD)~\cite{BD},  who discovered it in PF massive gravity supplemented with certain type 
 of non-linear interactions.  This analysis was generalized by Gabadadze and Gruzinov~\cite{gregoryandrei} with the same conclusion.  
 More recently, some important observations were made in~\cite{ghostcedric} and~\cite{ghostnicolis}. 
 In~\cite{ghostcedric} it was shown that in massive gravity recovery of the Einsteinian metric 
 at distances $r \ll r_{\star}$ can be understood as the result of the BD-type ghost exchange.  
 On the other hand, in~\cite{ghostnicolis} the presence  of the ghost was detected even at scales 
 $r \gg r_{\star}$. 
 
   In our analysis of phase transition from the degravitated vacuum in massive spin-2 theory, we encounter a time-like version of the above phenomenon.  
   Although in the decoupling limit of massive gravity  the cosmological constant degravitates, 
   during a phase transition the following happens.  Just like for space-localized sources, associated with a time-localized phase transition is a characteristic scale $t_{\star}$ --- the time-like analogue of $r_{\star}$. For early times, $t \ll t_{\star}$, the contribution from the St\"uckelberg field is suppressed relative to the
helicity-2 part, so that the metric responds to the transition in an almost Einsteinian way. However, the
transition excites the BD ghost-like mode, which can trigger an instability. Indeed, in certain cases we find
that, for $t \gg t_\star$, the St\"uckelberg field not only catches up with helicity-2, but actually overshoots
and grows unbounded exponentially in time.
  
     We should note that situation is different for generic resonance graviton. For instance, 
  it is known that such instabilities are absent in the DGP model. From the perspective of the helicity-0 graviton the difference with massive gravity is that
   in DGP the equation for the helicity-0 mode is not higher-derivative~\cite{ghostnicolis, 
  ghostcedric}. On the other hand, in DGP the cosmological constant is not degravitated either. 
  
  So a key question is whether there exists any sensible theory of the resonant graviton that combines 
  the absence of ghost instabilities  with degravitation.   We show that the requirement of degravitation, 
 at least in the decoupling limit of generic $m^2(\square)$ theories, puts a severe restriction 
 on the dispersion relation of the graviton, with the allowed range being bounded by 
 DGP resonance and massive gravity. Whether there is a non-empty  set of consistent theories within this interval  is an open question.  
  
 \subsection{Phenomenological Picture}
 
 Assuming that this set is not empty,  we then apply the $r_\star$-concept to theories with degravitating
cosmological constant, as described by the effective equation (\ref{central1}), and reach the following verdict.  

Any sensible theory of this sort must be part of a generally covariant theory of massive 
or resonance  graviton.  Equation~(\ref{central1}) must describe the effective behavior 
of the Einsteinian part of the metric only, once the St\"uckelberg fields have been integrated out. 
Since the contribution of extra polarizations to the full metric is suppressed on high-curvature backgrounds,~(\ref{central1}) can only be considered a good 
approximation for the dynamics of the full metric as long as $H \gg L^{-1}$.
In this regime the role of the extra polarizations is limited to a small contribution that takes care of the Bianchi
identity. Once the Hubble parameter drops to $\sim L^{-1}$, however, the extra 
polarizations become important, and~(\ref{central1}) can no longer be trusted. Having identified the applicability range of~(\ref{central1}), we then derive some
exact solutions describing degravitation dynamics for some examples  of filters.  
  
  Ideally one would like to speculate that, once the curvature of the Universe drops to $L^{-2}$ and the extra polarizations become relevant, 
 the Universe gracefully exits from the degravitation regime and is left with a small effective cosmological constant set by $L^{-2}$.
  Unfortunately,  we cannot make any rigorous statement  about this epoch, because
  both perturbative expansions, either in $1/L$ or $G_N$, break down. The question can only be answered by finding  
  exact solutions of the full theory. Of course we should stress that not every non-linearly completed theories of the resonant graviton
   will display degravitation. However, a key lesson is that the converse statement must be true: any degravitating theory must be a theory of a
   massive  or resonance graviton.    
  Such theories then should fall within the general category of large distance modified gravity theories  that can  be testable by  precision gravitational measurements~\cite{gd}.

\section{Inevitability of Extra Polarizations} \label{newdof}

We shall work with the following equation
\begin{equation}
\label{central}
\left (1 \, - \, {m^2(\square) \over \square}\right) \mathcal{G}_{\mu\nu} \, = \,  \frac{1}{2} T_{\mu\nu} \,,
\end{equation}
where $m^2(\Box) \propto L^{2(\alpha \, -\, 1) }\Box^{\alpha}$ with $\alpha < 1$ for $\Box L^2 \rightarrow 0$, while $m^2(\Box) \ll \Box$ for $\Box L^2\gg 1$.
An important lower bound on the parameter $\alpha$, following from unitarity~\cite{gd},  is $\alpha \geq 0$.
 This constraint implies that in the above theory, gravitons on a flat background must either have the dispersion 
 relation of a massive particle ($\alpha = 0$), or that of a resonance. The mass or width of the graviton is set by the filter 
 scale $L^{-1}$.

However, this is puzzling since, on the one hand, we know that in such a theory the graviton must propagate 
 three extra degrees of freedom. In the absence of ghosts, these are responsible for the vDVZ discontinuity at the linear level~\cite{vDVZ}. 
On the other hand,~(\ref{central}) continuously reduces to Einstein's theory in the limit $L \rightarrow \infty$, or equivalently $m^2(\square) \rightarrow 0$, 
 even at the linear level.

The resolution to the above puzzle is the following.
When expanded in small perturbations around the Minkowski background,~(\ref{central}) only describes the 
propagation of the helicity-2 degrees of freedom.
Below we denote this Einsteinian part of the small metric perturbations by $\tilde{h}_{\mu\nu}$. 
The full metric fluctuation $h_{\mu\nu}$, however, must contain 5
physical polarizations.  This fact also explains why~(\ref{central}) does not satisfy the Bianchi identity. 

To prove this, we show that~(\ref{linear})  is the equation satisfied by the two helicity-$\pm 2$ states 
 of the massive or resonance graviton, once the remaining 3 helicities have been integrated out. 
 Although this is done in~\cite{gd}, because of its importance for our purposes we will repeat the derivation in detail here.   

Our starting point is the most general ghost-free theory for a resonance graviton that involves all polarizations
\begin{equation}
\label{pf}
 \mathcal{E}^{\;\;\alpha\beta}_{\mu\nu} h_{\alpha\beta}\, 
\, - \, m^2(\Box) \left(h_{\mu\nu} \, - \, \eta_{\mu\nu} h\right) \,  = \, - T_{\mu\nu} \;.
\end{equation}
This is a consistent generalization of the Pauli-Fierz equation for a massive graviton~\cite{FP}, with 
the graviton mass $m^2$ promoted into a function of the d'Alembertian. 
As in Pauli-Fierz theory, the graviton satisfying~(\ref{pf}) propagates five degrees of freedom:
two spin-2, two spin-1 and one spin-0 helicities.

The next step consists in rewriting~(\ref{pf}) in a manifestly gauge-invariant form using the St\"uckelberg method.
That is, introducing a St\"uckelberg vector $A_\mu$, we express $h_{\mu\nu}$ in the following form 
\begin{equation}
\label{vector}
h_{\mu\nu} \, = \, \hat{h}_{\mu\nu} \, + \, \partial_{\mu}A_{\nu} \, + \, \partial_{\nu} A_{\mu} \,,
\end{equation}
with $A_{\mu}$ carrying the extra polarizations. In terms of $\hat{h}_{\mu\nu}$ and $A_{\mu}$,~(\ref{pf}) takes the form
\begin{equation}
\label{pfs}
 \mathcal{E}^{\;\;\alpha\beta}_{\mu\nu} \hat{h}_{\alpha\beta}\, 
\, - \, m^2(\Box) \left(\hat{h}_{\mu\nu} \, - \, \eta_{\mu\nu} \hat{h}  \, + \partial_{\mu}A_{\nu} \, + \, \partial_{\nu} A_{\mu} \, 
- \, 2 \eta_{\mu\nu} \partial^{\alpha}A_{\alpha}\right) \,  = \, - T_{\mu\nu} \,,
\end{equation}
and is now manifestly invariant under the gauge transformation 
\begin{equation}
\label{gauge}
\hat{h}_{\mu\nu} \, \rightarrow \hat{h}_{\mu\nu}  \, +  \, \partial_{\mu}\xi_{\nu} \, + \, \partial_{\nu} \xi_{\mu},
~~~A_{\mu} \, \rightarrow \, A_{\mu} \,  - \, \xi_{\mu} \,.
\end{equation}
Note that the Einstein tensor only supports the helicity-2 components --- it is unchanged under the replacement (\ref{vector})
and therefore gauge-invariant under (\ref{gauge}).

We now integrate out $A_{\mu}$ using its equation of motion, which follows from taking the divergence of~(\ref{pfs}):
\begin{equation}
\label{Aequ}
\partial^{\mu} F_{\mu\nu} \, = \, - \, \partial^{\mu}  \left(\hat{h}_{\mu\nu} \, - \, \eta_{\mu\nu} \hat{h} \right)\,,
\end{equation}
where $F_{\mu\nu} \, = \, \partial_{\mu}A_{\nu}  - \partial_{\nu}A_{\nu}$. 
Before solving for $A_{\mu}$, note that taking the divergence of~(\ref{Aequ})
yields the following constraint:
 \begin{equation}
\label{hconstraint}
 \partial^{\mu} \partial^{\nu}  \hat{h}_{\mu\nu} \, -  \, \Box \hat{h} \, = \, 0\; ,
\end{equation}
which is the statement that $\hat{h}_{\mu\nu}$ gives rise to a vanishing Ricci scalar. 
This implies that $\hat{h}_{\mu\nu}$ can be represented in the form 
\begin{equation}
\label{tilderep}
\hat{h}_{\mu\nu} \, = \, \tilde{h}_{\mu\nu} \, - \, \eta_{\mu\nu} {1\over 3} \Pi_{\alpha\beta}\tilde{h}^{\alpha\beta}\,,
\end{equation}
where $\Pi_{\alpha\beta}\, = \, \eta_{\alpha\beta} \, - \, \partial_{\alpha}\partial_{\beta}/ \Box$ 
is the transverse projector.  Thus $\tilde{h}_{\mu\nu}$ carries two degrees of freedom.
Moreover, since the last term in (\ref{tilderep}) is gauge-invariant, $\tilde{h}_{\mu\nu}$ shifts in the same way as $\hat{h}_{\mu\nu}$
under the gauge transformations~(\ref{gauge}).

Coming back to~(\ref{Aequ}), we can solve for $A_{\mu}$:
\begin{equation}
\label{asolution}
A_{\nu} \, = \, - \, {1 \over \Box} \partial^{\mu}  \left(\hat{h}_{\mu\nu} \, - \, \eta_{\mu\nu} \hat{h} \right) \, - \, \partial_{\nu} \Theta\,, 
\end{equation}
where $\Theta$ is an arbitrary function. Gauge invariance demands that under the transformations~(\ref{gauge}), $\Theta$ shifts in the following way
\begin{equation}
\label{thetagauge}
\Theta \, \rightarrow \, \Theta \, + \, {1 \over \Box} \partial_{\alpha}\xi^{\alpha}\,.
\end{equation}
Substituting~(\ref{asolution}) back into~(\ref{pfs}), and expressing  
$\hat{h}_{\mu\nu}$ in terms of  $\tilde{h}_{\mu\nu}$ using~(\ref{tilderep}), we arrive at the following
effective  equation for $\tilde{h}_{\mu\nu}$
\begin{equation}
\label{pftilde}
\left(1\,  - \, {m^2(\Box) \over \Box} \right )  \mathcal{E}^{\;\; \alpha\beta}_{\mu\nu} \tilde{h}_{\alpha\beta}\, -  \, \Box \Pi_{\mu\nu} \left [ \tilde{h}  \, - \,  2 \left( 1\,  - \, {\Box \over  3 m^2} \right ) \Pi_{\alpha\beta}\tilde{h}^{\alpha\beta}\,  +  \, 2 \Theta \right ] \,  = \, - T_{\mu\nu} \,.
\end{equation}
The expression in square brakets can be set to zero by appropriate choice of $\Theta$ or, equivalently,
by making a gauge transformation in $\tilde{h}$ and $\Theta$ with the gauge parameter $\xi_{\mu}$ satisfying 
\begin{equation}
\label{xifix}
\left(1\,  + \, {1 \over \Box} \right )  \partial_{\mu}\xi^{\mu}\, = \, 
 \left( 1\,  - \, {\Box \over  3 m^2} \right ) \Pi_{\alpha\beta}\tilde{h}^{\alpha\beta}.
\end{equation}
The resulting equation is~(\ref{linear}), which proves our point. 

Because $\tilde{h}_{\mu\nu}$ propagates only two degrees of freedom, the one particle exchange amplitude between two 
conserved sources, $T_{\mu\nu}$ and  $T_{\mu\nu}'$, mediated by $\tilde{h}$ is
\begin{equation}
\label{amplitudepart}
({\rm Amplitude})_{\tilde{h}} \, \propto \, {T'_{\mu\nu}T^{\mu\nu} \, - {1 \over 2} T\; T'
\over \Box \,  - \, m^2(\Box)}\,,
\end{equation}
which continuously recovers the massless graviton result in the limit $m^2 \rightarrow 0$. 
This fact, however, {\it does not} avoid the well known vDVZ discontinuity since~(\ref{amplitudepart})
is only part of the full one-particle exchange amplitude.  This becomes immediately clear once we 
notice that the metric excitation that couples to the conserved source is $\hat{h}_{\mu\nu}$ (or equivalently  $h_{\mu\nu}$), which
depends on $\tilde{h}_{\mu\nu}$ through~(\ref{tilderep}).  The full physical amplitude 
is generated by the latter combination of $\tilde{h}_{\mu\nu}$ and is equal to 
\begin{equation}
\label{amplitudepart2}
({\rm Amplitude})_{h} \, \propto \, {T'_{\mu\nu}T^{\mu\nu} \, - {1 \over 3} T\; T'
\over \Box \, - \, m^2(\Box)}\,,
\end{equation}
which clearly exhibits the vDVZ-type discontinuity in the $m^2(\Box)\, \rightarrow \, 0$ limit.

\section{The Strong Coupling of the Longitudinal  Gravitons}

We have proved that in the considered class of theories the graviton must contain extra 
{\it longitudinal} polarizations. 
In this section we argue that the longitudinal states must be strongly coupled. 
The strong coupling of the extra states for massive gravity ($\alpha \, = \,  0$) 
 and for the DGP model ($\alpha \, = \, 1/2$) was discovered in~\cite{ddgv} and generalized  
for arbitrary $\alpha$ in~\cite{gd}. We shall utilize the analysis presented there.

Consider a generally covariant theory which, in the linearized approximation, 
reduces to~(\ref{pf}). The propagator of the graviton is 
\beq
\label{prop}
D_{\mu\nu ;\alpha\beta}\,=\, 
\left(
{1\over 2} \,\tilde\eta_{\mu\alpha} \tilde \eta_{\nu\beta}+
{1\over 2} \, \tilde\eta_{\mu\beta}  \tilde\eta_{\nu\alpha}-   
{1\over 3} \, \tilde\eta_{\mu\nu} \tilde \eta_{\alpha\beta}\right)\frac{1}{
\Box\, - \, m^2(\Box)}\,,
\label{5D}
\eeq  
where 
\begin{equation}
\label{etatilde}
\tilde\eta_{\mu\nu}=\eta_{\mu\nu}\, - \,  {\partial_\mu \partial_\nu \over m^2(\Box)}
\, = \, \eta_{\mu\nu}\, - \, L^{2(\alpha\, - \,1)} {\partial_\mu \partial_\nu  \over \Box^{\alpha}}\,.
\end{equation}

The culprits for the strong coupling phenomenon are the terms singular in $L^{-1}$. 
These  terms arise from the additional helicity-0 mode of the resonance (or massive) graviton. In the St\"uckelberg language expressed 
in~(\ref{vector}), this helicity-0 polarization is shared between $A_{\mu}$ and the trace of $\hat{h}_{\mu\nu}$.  
If we denote the canonically normalized  longitudinal  polarization by  $\chi$ and ignore the helicity-1 component, the full metric fluctuation
can be represented as 
\begin{equation}
\label{handchi}
h_{\mu\nu} \, = \, \tilde{h}_{\mu\nu} \,  + \, {1\over 6} \eta_{\mu\nu}\chi \, + \,
 {L^{2(\alpha\, - \,1)} \over 3}  {\partial_\mu \partial_\nu  \over \Box^{\alpha}}\chi\,,
\end{equation}
where $\tilde{h}_{\mu\nu}$ is the same as in~(\ref{tilderep}) and~(\ref{pftilde}).

The helicity-0 state is responsible for the extra attraction that provides the factor $1/3$ in the one graviton exchange amplitude (\ref{amplitudepart2}), 
as opposed to $1/2$ in standard gravity (\ref{amplitudepart}).  This is the essence of the
vDVZ discontinuity. The strong coupling of the longitudinal helicity-0  graviton, however, 
recovers continuity at the non-linear level. 

The net effect of this strong coupling is that near gravitating sources, where curvature effects are  
important, the relative contribution from the helicity-0 component gets suppressed. 
In a gauge-invariant language,  an observer measuring  the 
physical effect of the metric fluctuation $h_{\mu\nu}$ by a probe conserved energy momentum tensor $t_{\mu\nu}$ will mostly be exposed to the Einsteinian helicity-2
part:
\begin{equation}
\label{probing}
\int d^4 x \, t^{\mu\nu}h_{\mu\nu} \, \simeq \, \int d^4x \, t^{\mu\nu}\tilde{h}_{\mu\nu}\,.
\end{equation} 
As mentioned in the introduction, we refer to this effect as the $r_{\star}$-phenomenon. 
Due to strong coupling effects, a gravitating sources of mass $M$ acquires, on top of the usual 
Schwarzschild radius $r_g \equiv 2G_N \, M$, a second physical radius which we call $r_\star$.  
For the longitudinal graviton $\chi$, the latter radius plays the role somewhat similar to  the one played by  
$r_g$ for the transverse graviton  $\tilde{h}_{\mu\nu}$. Namely, due to the strong coupling
at $r=r_\star$, the non-linear self interactions of $\chi$ become  important, and the perturbation expansion in $G_N$ breaks down. 

The concept of $r_\star$ plays a central role in understanding the nature of gravitating sources.
Consider a localized static source, $T_{\mu\nu} \, = \, \delta_{\mu}^0\delta_{\nu}^0 M\delta(r)$, with Schwarzschild radius $r_g$.  
Very far away from the source the linear approximation should be valid. To linear order in $G_N$, the metric 
 created by this source can be calculated in the one graviton exchange approximation, with the result
\begin{equation}
\label{metricofM}
h_{\mu\nu}  \, = \,- {\delta_{\mu}^0\delta_{\nu}^0\, - {1 \over 3} \left(\eta_{\mu\nu}\, - \, L^{2(\alpha\, - \,1)} {\partial_\mu \partial_\nu  \over \Box^{\alpha}}\right)
\over \Box \,  - \, L^{2(1-\alpha)} \Box^{\alpha}} {r_g \over 2} \delta(r)\,.
\end{equation}
The term in the numerator that is singular in $1/L$ vanishes when convoluted with any conserved test
source $T'_{\mu\nu}$, in accordance with (\ref{amplitudepart}). Hence, at distances smaller than $L$, 
the metric has the usual $r_g/ r$ form, but with a {\it wrong} (scalar-tensor type) tensorial structure, 
manifesting the vDVZ discontinuity. However, if nonlinear interactions are introduced, 
the term singular in $1/L$ no longer vanishes and, in fact, diminishes the linear effects as a result of the strong coupling. 
The strong coupling scale was estimated in \cite{gd} by  generalizing the analysis of \cite{ddgv, dgz}. 
Power counting shows that  the leading singularity in $L$ comes from the trilinear 
interaction of the longitudinal gravitons and is of order $L^{4(1-\alpha)}$. This vertex has a momentum dependence of the form
\begin{equation}
\label{vertex}
      (Lk)^{4(1-\alpha)}k^2 \,.
\end{equation}
The  scale $r_\star$ then corresponds to the distance from the source at which the contribution 
from the above trilinear vertex becomes as important as the linear contribution given by~(\ref{metricofM}). This gives
 \begin{equation}
\label{rstar}
r_\star \, = \, \left(L^{4(1-\alpha)}r_g\right)^{{1 \over 1 + 4(1-\alpha)}} \; .
\end{equation} 

For distances $r \ll r_\star$, the correction to the Einstein metric from the longitudinal gravitons 
is suppressed by powers of $r_\star$. The leading behavior can be fixed by two requirements. 
First,  $\chi(r)$ must become of order $r_g/r_\star$ at $r=r_\star$ in order to match the linear regime~(\ref{metricofM})  
outside the $r_\star$-sphere. Second, it must be possible to approximate the solution inside $r_\star$  
by an analytic  series in $L^{2(\alpha-1)}$. These two requirements therefore fix the leading behavior to
\begin{equation}
\label{correction}
\chi \left(r \ll r_\star\right) \, \sim \,  {r_g \over r_\star} \left ({r \over r_\star}\right )^{{3 \over 2} - 2 \alpha} \,,
\end{equation} 
which results in the following relative correction to the gravitational potential 
\begin{equation}
\label{relativecor}
\delta \, \sim \, \left ({r \over L} \right )^{2(1-\alpha)} \sqrt{r \over r_g} \,.
\end{equation}
Such corrections are potentially measurable with upcoming high-precision experiments probing
gravity in the solar system~\cite{dgz, gd}, in particular Lunar Laser Ranging experiments~\cite{lunar}.  

\subsection{Extra States and Bianchi Identity} 

 Next we turn to the issue of the Bianchi identity. 
Already the fact that $h_{\mu\nu}$ propagates more than two degrees of freedom tells us that 
 an extreme care must be taken in working with~(\ref{central}).  
Because $\Box$ for a general background does not commute with $\nabla_{\mu}$, at first glance the Bianchi identity 
is either not satisfied or implies an additional constraint on
 the metric. Indeed, acting with the covariant derivative on~(\ref{central}) and using the Bianchi identity, 
 $\nabla^{\mu} \mathcal{G}_{\mu\nu} \, = \, 0$, we get 
\begin{equation}
\label{bianchi}
L^{2(\alpha \, -\, 1)}\left [ \nabla^{\mu}, \nabla^{2(\alpha\, - \, 1)} \right ]  \mathcal{G}_{\mu\nu} \, = \,
\nabla^{\mu} \, T_{\mu\nu}\,. 
\end{equation}
Naively, one would assume that the right hand side should vanish by covariant conservation 
of the source.  And, since $\left [ \nabla^{\mu}, \nabla^{2(\alpha\, - \, 1)} \right ] \,   \neq \, 0$ for a generic metric, 
one would conclude that~(\ref{bianchi}) necessarily implies an additional constraint on the metric. 
But this turns out to be false because the right hand side is in fact non-zero.  Although it is certainly true that 
$T_{\mu\nu}$ is covariantly conserved, it is conserved with respect to the full metric  
\begin{equation}
\label{fullmetric}
g_{\mu\nu} \, = \, \eta_{\mu\nu} \, + \, h_{\mu\nu} \, =  \eta_{\mu\nu} \, 
+ \, \tilde{h}_{\mu\nu} \,  + \, {1\over 6} \eta_{\mu\nu}\chi \, + \, {1 \over 3} 
 L^{2(\alpha\, - \,1)} {\partial_\mu \partial_\nu  \over \Box^{\alpha}}\chi \, + \, (\rm higher~order ~corrections) \,.
\end{equation}
Meanwhile, as shown above,~(\ref{central}) does not include the full metric, but only its transverse part: 
\begin{equation}
\label{partmetric}
\tilde{g}_{\mu\nu} \, = \, \eta_{\mu\nu} \, + \, \tilde{h}_{\mu\nu} \, \, + \, 
 \, (\rm higher~order~corrections)\,. 
\end{equation}
Since $\mathcal{G}_{\mu\nu}$ is the Einstein tensor with respect to  $\tilde{g}_{\mu\nu}$, as opposed to $g_{\mu\nu}$, it is covariantly
conserved only with respect  to the former metric
\begin{equation}
\label{nabla}
\tilde{\nabla}^{\mu} \mathcal{G}_{\mu\nu} \, = \, 0 \,,
\end{equation}
where $\tilde{\nabla}$ is the covariant derivative operator associated with $\tilde{g}$.
With respect to the same metric, however, the source is {\it not} conserved:
\begin{equation}
\label{source}
\tilde{\nabla}^{\mu}T_{\mu\nu} \, \neq \, 0\,.
\end{equation}
In other words, when acting with $\tilde{\nabla}^{\mu}$ on~(\ref{central}) we cannot simply
assume that the divergence of the right hand side vanishes.  

In fact it is easy to estimate the range of validity of~(\ref{central}).  Consider a background 
created by a source of size $R$ localized within its own $r_\star$, {\it i.e.}, $R \ll r_\star$. 
As we have seen in the previous section, 
the metric within such a source is almost Einsteinian, with small corrections given by~(\ref{correction}) 
and~(\ref{relativecor}) coming from the longitudinal polarization $\chi$. Thus, for $r \ll r_\star$,
 \begin{equation}
\label{metricplus}
g_{\mu\nu} \, = \, \tilde{g}_{\mu\nu} \, + \, \mathcal{O} \left( \left ({r \over r_\star}\right )^{{5 \over 2} - 2 \alpha} \right )
\, = \,  \, \tilde{g}_{\mu\nu} \, + \, \mathcal{O} \left( \left ({r \over L} \right )^{2(1-\alpha)} \sqrt{{r \over r_g}} \right ) \,,
\end{equation}
and, as a result,
 \begin{equation}
\label{source2}
\tilde{\nabla}^{\mu} \, T_{\mu\nu} \, = \,\mathcal{O} \left( \left ({r \over L} \right )^{2(1-\alpha)} \right ) \,.
\end{equation}
Thus we see that both sides~(\ref{bianchi})  are of order $(r / L)^{2(1-\alpha)}$, in accordance with the Bianchi identity.
  
The above analysis also clarifies the range of scales and sources for which the solutions to~(\ref{central}) 
are close to those from Einstein's theory. For any given gravitating source, the solutions 
to~(\ref{central}) are a good approximation to their counterpart in Einstein gravity on scales $r \ll r_\star$, up to small corrections given by~(\ref{relativecor}). 
For such sources and distances the contribution of the extra states can be ignored, 
and the full metric is well approximated by the solutions of~(\ref{central}).  At distances 
around $r \sim r_\star$, however, the contribution from extra states becomes important  and must not be ignored. 
At such distances, the solutions to~(\ref{central}) must be supplemented by contributions from longitudinal graviton dynamics. 

\section{De-Electrifying  the Vacuum}

 Before turning to gravity, we wish to highlight some of the key ideas with a much simpler system of 
 the abelian spin-1 gauge theory.  Starting with the massless Maxwell's electrodynamics, we 
define the electric analogue of the cosmological constant problem. Essentially, that a uniform constant   
charge density produces a linearly growing electric field.  Then we show that this problem can be cured by 
introducing a small Proca (or Higgs) mass for the photon. This mass screens any background charge
density and `de-electrifies'  the vacuum. By integrating out the Goldstone-St\"ukelberg field, we show 
that de-electrification is equivalent to filtering.  We then add non-linear interactions of the Proca photon 
and show that there is a simple analogue of the helicity-0 graviton strong coupling, as well as the $r_{\star}$ phenomenon. 
 
\subsection{Filtering Vectors}

 In Maxwell's electrodynamics, an infinite wavelength source is not filtered and produces a 
 non-zero electric field. This is exactly as in massless gravity where even a uniform source 
 results in non-zero curvature. In both cases, there is no filtering effect, basically because there is no mass scale available to define an intrinsic filter. 
 Hence, in Maxwell's electrodynamics this is the analogue of the cosmological constant problem. 
 
The maximally symmetric uniform  source, which is the closest possible electromagnetic analogue of the cosmological term,  is of the form
\begin{equation}
\label{chargecc}
J_\alpha \equiv \delta_\alpha^0 J_0\,,
\end{equation}
with $J_0$ a constant.  This source represents a uniformly distributed charge density.  
The essence of the problem is that  such a charge produces a linearly growing electric field 
$E_j = -x_j J_0/3$, corresponding to the static Maxwell-field 
\begin{equation}
\label{statica}
\Delta \tilde{A}_{0} = -J_0\,.
\end{equation}

There are the two seemingly different  strategies to solve this problem. 
One possibility is to  modify Maxwell's electrodynamics in such a way that  the photon is no longer sourced by constant 
(or very large wavelength) sources. That is,  by invoking a filter we can modify Maxwell's equation in the following way 
\begin{equation}
\label{maxfilter}
\left (1 \, -\, {m^2 \over \square} \right )  \partial^\beta F_{\beta\alpha} =
- J_{\alpha}\,.
\; , 
\end{equation}
It is now obvious that uniform sources of the form~(\ref{chargecc}) no longer produce an electric field and instead get screened (or `de-electified').  

The second option is to give a small mass  ($m$) to the photon, thereby promoting it from a
Maxwell field $\tilde{A}_{\mu}$ to a Proca field  $A_{\mu}$. In the presence of a mass term, the vacuum becomes superconducting, 
and all charges are screened at distances much larger than $m^{-1}$.  

However, these two methods are precisely equivalent, as we now explicitly show.
 In both cases, the culprit for screening the background charge is the 
 Goldstone-St\"uckelberg field. The filter~(\ref{maxfilter}) is just an effective description of this
 dynamics. 

Consider, therefore, a Proca field of mass $m$ coupled to an external source:
\begin{equation}
\label{lpm}
{\cal L} 
=
-\frac{1}{4} F^2 - \frac{1}{2} m^2 A^2 + J_\mu A^\mu
\; .
\end{equation}
In this case, even if $J$ is conserved, the explicit mass term breaks gauge invariance. As before, the
mass and interaction terms can be made gauge-invariant through the St\"uckelberg field, as in~(\ref{stuckelphoton}).  
The main difference with Maxwell's theory is that the mass term in~(\ref{lpm}) makes $\phi$ dependent on $\tilde{A}$: 
\begin{equation}
\phi 
=
-m\frac{\partial^\alpha}{\Box}\left(\tilde{A}_\alpha - \frac{1}{m^2} J_\alpha\right)
\; .
\end{equation}
Substituting this into the Proca equation for $A_{\mu}$, we recover (\ref{maxfilter})
for $\tilde{A}_{\mu}$. As usual, the longitudinal projection guarantees that~\footnote{The operators
$\theta_{\mu\nu} \equiv \partial_\mu \partial_\nu/\Box$ and $\Pi_{\mu\nu} \equiv \eta_{\mu\nu} - \theta_{\mu\nu}$
project on longitudinal and transversal contributions, respectively. They fulfill the following identities:
$\theta_{\mu\sigma}\theta^{\sigma\nu} = \theta_\mu^{\; \nu}$, $\theta_{\mu\sigma}\Pi^{\sigma\nu} = 0$ and
$\Pi_{\mu\sigma}\Pi^{\sigma\nu} = \Pi_\mu^{\; \nu}$. Thus, 
$\tilde{A}^{||}_{\; \alpha} \equiv \partial_\alpha \Psi (\tilde{A}) =\theta_{\alpha\beta} \tilde{A}^\beta$ and 
$\tilde{A}^{\perp} (\tilde{A}) = \Pi_{\; \alpha\beta} \tilde{A}^\beta$.
Note that $\tilde{A}^{||}$ is a derived field, the basic fields are $\tilde{A}^{\perp}$ and the 
scalar potential $\Psi \equiv \partial\cdot \tilde{A}/\Box$
that generates the longitudinal polarization.}
\begin{equation}
A^{||}_{\; \beta} 
=
-\frac{\partial_\alpha\partial_\beta}{\Box} \left(\tilde{A}^\beta -\frac{1}{m^2} J^\beta \right)
\end{equation}
can be eliminated from the dynamical problem. Thus, in Proca theory, screening of the background charge for the Maxwellian 
component $\tilde{A}_{\mu}$ can be reformulated equivalently in terms of filtering.  

In this example, the St\"uckelberg-field fulfills two purposes. First, it allows a gauge-invariant extension of the original dynamical variable, thereby
making the theory manifestly gauge-invariant and ensuring only physical
interactions. Second, by integrating it out, one is left automatically with the true Maxwellian degrees of freedom.  

Let us study the effect of the Proca filter, again for the uniform source $J_\alpha \equiv \delta_\alpha^0 J_0$.
The equation of motion for the components of $\tilde{A}$ is, of course,
just the Klein-Gordon equation with the specified source. The solution of interest is given by
\bea
\label{dechargeproca}
\nonumber
\tilde{A}_{0} &= &\frac{J_0}{m^2} \left(1-\cos mt\right)\;;\\
\tilde{A}_{j} &=& \frac{x_j}{3} \dot{\tilde{A}}_{0}\,,  
\eea
describing coherent oscillations about the vacuum with zero electric field. 
The amplitude of the oscillations is of course arbitrary, but we have chosen it such that for $t \ll m^{-1}$,
~(\ref{dechargeproca}) smoothly reduces to the massless solution with linear electric field $E_j = \, -x_jJ_0/3$. 
In the presence of dissipation, these oscillations will damp down, and the 
system will settle to the vacuum with no electric field, {\it i.e.},  it will get de-electrified.  
And, as we have seen, this de-electrification is completely equivalent to 
filtering.    

The next question is what happens when non-linearities are included? 
In short, non-linearities do not destroy de-electrification --- in the 
the presence of a uniform charge density, the vacuum continues to have zero electric 
field. However, there is an interesting new twist. In the non-linear case, as we will see, the linear filtering studied above is not always the dominant effect in de-electrifying the vacuum. 
 
Thus consider the following toy non-linear theory for the Proca field, obtained by adding a quartic interaction:
\be
\label{rw}
{\cal L}
=
-\frac{1}{4} F^2 - \frac{1}{2} m^2 A^2 + J_\mu A^\mu
- \frac{\lambda}{4} \left(A_\mu A^\mu\right)^2
\; .
\ee 
Evidently, non-linearities do not spoil de-electrification. For the constant source~(\ref{chargecc}), 
there is a static solution $A_{\mu}\, = \, \delta_{\mu}^0\, a$, where $a$ satisfies the cubic equation
\be
\label{roots}
  m^2 \, a  \, + \, \lambda \, a^3 \, = \,  J_0\,.
\ee
Starting from arbitrary initial conditions, in the presence of dissipation the system will 
settle to this final state. For appropriately small $\lambda$,  this de-electrification 
is understood as a filtering effect, which can be shown explicitly by integrating out the 
St\"uckelberg field  order by order in $\lambda$ and finding that the resulting theory 
is a small perturbation to the effective filter. 
For large $\lambda$, however, we see from~(\ref{roots}) that de-electrification proceeds through the non-linear term.
Below we will discuss how this relates to the $r_\star$-phenomenon in the St\"uckelberg language.

\subsection{Decoupling of St\"uckelberg and $r_\star$-Effect  for Proca Field}

Consider again a massive Proca field with quartic self-interaction, as given by~(\ref{rw}). Once again
we can make $A_\mu$ gauge-invariant by introducing a St\"uckelberg-field, as in~(\ref{stuckelphoton}).
As $m\rightarrow 0$, the strongest singularity comes from the $(\partial\phi)^4$
contribution in the self-interaction. Hence, the characteristic scale for the strong coupling regime of the 
St\"uckelberg field is
\be
\Lambda_{\rm strong} \,  \equiv \,  \frac{m}{\lambda^{1/4}}\,. 
\ee

Thus let us take the decoupling limit $m \, \rightarrow \, 0$ keeping  $\lambda/m^4$ fixed.
That is, we keep the strong coupling scale $\Lambda_{\rm strong}$  of the Goldstone-St\"ckelberg field fixed.   
In this limit, the equations for the massless photon $\tilde{A}_{\mu}$ and the St\"uckelberg field $\phi$ decouple:
\begin{eqnarray}
\nonumber
&& \partial^\alpha F_{\alpha \beta} = -J^\perp_{\; \beta}
\; , \\
&& \left[
1+ \frac{\lambda}{m^4}\left(\partial \phi\right)^2
\right]\partial_\beta \phi
=
m^{-1} \, J^{||}_{\; \beta}
\; .
\label{ans2}
\end{eqnarray}
For the source let us take $J_\alpha = m\;  \delta_\alpha^{\; 0} \; t f(r)$, corresponding to $J^{||}_{\; \alpha} = (\partial_\alpha/\Box) m f(r)$ and
$J^{\perp}_{\; \alpha} = - m \; \delta_\alpha^{\; 0} \; t f(r) - J^{||}_{\; \alpha}$. 

In order to study the interpolating behavior of the St\"uckelberg-field in this decoupling limit,  
it suffices to solve the static problem. The first of~(\ref{ans2}) is satisfied by 
$\tilde{A}_{\; 0} = (m/\Delta) t f(r)$ and $\tilde{A}_{\;i} \, = \, 0$.
By spherical symmetry of the source term, the second of~(\ref{ans2}) reduces to
\begin{equation}
\label{red}
\frac{1}{\Lambda_{\rm strong}^4} \left(\partial_r \phi\right)^3 +  \partial_r \phi
=
 \partial_r \left(\frac{1}{\Delta} f(r)\right)
 \; .
 \end{equation}
It is obvious that when nonlinearities are negligible we have $\phi = (1/\Delta)f(r)$.
In the opposite limit where the nonlinear term dominates, 
 the solution is $\partial_r \phi\, = \, \Lambda_{\rm strong}^{4/3} \, [\partial_r (1/\Delta) f(r)]^{1/3}$. 

\begin{figure}[ht]
\includegraphics[width=120mm]{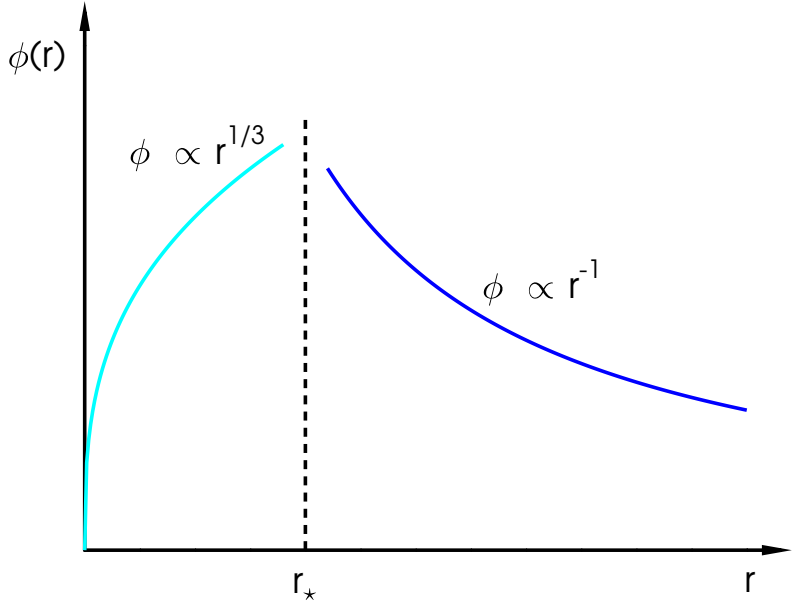}
\centering
\caption{The weak coupling ($r>r_\star$) and strong coupling ($r<r_\star$) solution of (\ref{red}) for the St\"uckelberg-field
in the case of a point charge at $r=0$.  The change in the behavior at $r_\star$ is analogous to the gravity case.}
\label{stueckel}
\end{figure}

For definiteness,  consider a point charge at $r=0$, corresponding to $f(r)=4\pi Q\; \delta(r)$. 
In this case, we have
\begin{equation}
\label{phismall}
\phi (r) \, \simeq \,  3 (Q \Lambda_{\rm strong}^4)^{1/3} r^{1/3}~~~~~~ {\rm for}~~r \ll r_\star\,,
\end{equation}
and 
\begin{equation}
\label{philarge}
\phi (r) \, \simeq \,  {Q \over r}~~~~~{\rm for} ~~~r \gg r_\star\,,   
\end{equation}
where the scale $r_\star$ is defined as 
\begin{equation}
\label{rstarvector}
r_\star\equiv   {Q^{1/2} \over \Lambda_{\rm strong}} \,.
\end{equation}
This solution is sketched in Fig.~\ref{stueckel}. 

This interpolating behavior highlights some similarities and differences between the 
spin-1 and spin-2 cases.  In both cases,  the strong coupling effect of the St\"uckelberg 
field leads to an $r_\star$-phenomenon.  Namely, due to non-linearities, 
the St\"uckelberg field is strongly suppressed relative to the expected $1/r$ linear behavior.
Unlike the spin-2 case, however,  such a suppression has limited importance, since for spin-1 the  St\"uckelberg field does not couple to conserved probe sources. 
For massive spin-2, on the other hand, the helicity-0  St\"uckelberg field couples to conserved sources, and the $r_\star$-phenomenon is crucial for the phenomenological viability of the theory.  
In the next section we apply a similar analysis to massive spin-2.
 
\section{Spin-2 Case: Linearized Filtering and Role of the Graviton Mass}
\label{spin2}

For simplicity, we first study the spin-2 case at the linearized level.  First we show that the cosmological constant problem is present even at the level
of the linearized massless theory. We then argue that this problem can be resolved by allowing for a small graviton mass, which is equivalent to introducing a filter.
In analogy with the vanishing electric field solutions in the spin-1 case, here we find solutions describing flat space even in the presence of vacuum energy. Moreover we find
oscillatory solutions around this new vacuum. The close parallel with Proca theory will be apparent at every step.

Consider the linearized equation for a massless spin-2 field that is sourced by a cosmological term,  
\begin{equation}
\label{linearcc}
 \mathcal{E}^{\;\; \alpha\beta}_{\mu\nu} \tilde{h}_{\alpha\beta}\,
 = \Lambda \eta_{\mu\nu}\,,
\end{equation}
where $\mathcal{E}^{\;\; \alpha\beta}_{\mu\nu}$ is defined in~(\ref{einstein})~\footnote{Consistency demands that terms of the form $\Lambda h_{\mu\nu}$ are
not included to this order.}.
This has a number of gauge-equivalent solutions corresponding to different 
slicings of de Sitter space-time. For example, 
\be
\label{linds}
\tilde{h}_{00} \, = \, 0,~~ \tilde{h}_{0i} \, = \, 0,~~
\tilde{h}_{ij} \, = \, {\Lambda \over 6} (t^2 \delta_{ij} \, + \, x_ix_j)
\ee
is the linearized de Sitter metric in closed FRW slicings. Similarly,  
\be
\label{statds}
\tilde{h}_{00} \, = \,  {\Lambda \over 6}  x_ix^i,~~ \tilde{h}_{0i} \, = \, 0,~~
\tilde{h}_{ij} \, = \,  {\Lambda \over 6} \, x_ix_j 
\ee
describes the static patch of de Sitter space.

Next we investigate how the cosmological constant gets degravitated due to a finite graviton mass. Our starting point is the linearized Fiertz-Pauli equation~(\ref{pf}):
\be
- \mathcal{E}^{\;\;\alpha\beta}_{\mu\nu} h_{\alpha\beta} + m^2\left(h_{\mu\nu}-\eta_{\mu\nu}h\right) = -\Lambda\eta_{\mu\nu}\,.
\label{linear2}
\ee
As shown above, in Einstein gravity ($m=0$) the cosmological
term acts as a tadpole and generates an instability of flat space towards de Sitter spacetime. With a finite graviton mass, however, flat space
becomes a solution
\be
h_{\mu\nu} =  \frac{\Lambda}{3m^2}\eta_{\mu\nu}\,.
\label{flatshift}
\ee
In other words, the $\Lambda$-source has the effect of shifting the usual $h_{\mu\nu} = 0$ Minkowski vacuum 
to a new one given by~(\ref{flatshift}), which also describes flat space.
This is the analogue of the massive vector solution $A_0 = J_0/m^2$ obtained in the previous section.

To underscore the high-pass filter nature of massive gravity, let us decompose the 5 polarizations 
of the metric $h_{\mu\nu}$, as we did in Sec.~\ref{newdof}, and study the evolution of the helicity-2 component $\tilde{h}_{\mu\nu}$. 
This satisfies~(\ref{linear}), which obtains after integrating out all other degrees of freedom:
\be
 \left ( 1 \, - \, {m^2\over \Box} \right ) \mathcal{E}^{\;\; \alpha\beta}_{\mu\nu} \tilde{h}_{\alpha\beta}
\, = \,  \Lambda \eta_{\mu\nu} \,.
\ee
To illustrate how degravitation takes place dynamically, we seek solutions which at early times $t\ll m^{-1}$ reproduce de Sitter expansion, 
as in Einstein gravity, while at late times  $t\gg m^{-1}$ oscillate around the flat space solution~(\ref{flatshift}).

The analysis is most easily done in de Donder gauge, $\partial^\mu\left(\tilde{h}_{\mu\nu} - \eta_{\mu\nu} \tilde{h}/2\right) =0$, 
in which the equation of motion simplifies to a massive wave equation
\be
\left(-\Box + m^2\right) \left(\tilde{h}_{\mu\nu}-\frac{1}{2}\eta_{\mu\nu}\tilde{h}\right)  = -\Lambda\eta_{\mu\nu}\,.
\label{dedonder}
\ee
The solution of interest, consistent with the gauge choice and our initial conditions, is then
\bea
\nonumber
\tilde{h}_{00} &=& -\frac{\Lambda}{m^2}\left(1-\cos(mt)\right)\;; \\
\nonumber
\tilde{h}_{0i} &=&  \frac{\Lambda}{3m} \sin(mt)\, x_i \;; \\
\tilde{h}_{ij} &=&  \frac{\Lambda}{m^2}\left(1-\cos (mt)\right)\delta_{ij} + \frac{\Lambda}{6} \cos(mt) \epsilon_{ij}\,,
\label{osc}
\eea
where $\epsilon_{ij} \equiv x_ix_j$ for $i\neq j$ and 0 otherwise. To see that this is approximately de Sitter at early times, note that for $mt\ll 1$ we get
\be
\tilde{h}_{00} \approx  - \frac{1}{2} \Lambda t^2\;; \qquad \tilde{h}_{0i} \approx  \frac{1}{3}\Lambda t x_i\;; \qquad
\tilde{h}_{ij} \approx  \frac{1}{2} \Lambda t^2 \delta_{ij} + \frac{\Lambda}{6} \epsilon_{ij}\,,
\ee
which can be recast through a small diffeomorphism into the more cosmologically-friendly form: 
\bea
\nonumber
ds^2 &=& -dt^2 + \left(1+\frac{\Lambda t^2}{3}\right)\delta_{ij}dx^idx^j + \frac{\Lambda}{3} x_ix_idx^idx^j \\
&\approx & -dt^2 + \cosh \sqrt{\frac{\Lambda}{3}}t \left(\frac{dr^2}{1-\Lambda r^2/3}+r^2d\Omega^2\right)\,.
\eea
This is of course just de Sitter in closed FRW slicing. 

Hence, as advertised,~(\ref{osc}) approximates de Sitter at early times and subsequently oscillates about the new flat space vacuum. If we turn on interactions with massless fields, these oscillations will be damped by particle production, and the geometry will eventually settle into the flat space vacuum~(\ref{flatshift}). 

It is easy to reconstruct how the degravitation story of $\tilde{h}_{\mu\nu}$ carries over 
to the gauge-invariant metric perturbation $h_{\mu\nu}$.   
Without loss of generality, we can set the helicity-1 component to zero.   
Then, in terms of a canonically-normalized helicity-2 state $\tilde{h}_{\mu\nu}$ and a helicity-0 state $\chi$, the gauge-invariant perturbation
can be written as~(\ref{handchi})
\begin{equation}
\label{hdecomp}
h_{\mu\nu} \, = \, \tilde{h}_{\mu\nu} \,  + \, {1\over 6} \eta_{\mu\nu}\chi \, + \,
 {1 \over 3} {\partial_\mu \partial_\nu  \over m^2}\chi\,.
\end{equation}
Canonical normalization refers to the fact that the kinetic terms for $\tilde{h}_{\mu\nu}$ and $\chi$ are diagonal in this basis.
Substituting (\ref{hdecomp}) into (\ref{linear2}) we get 
 \be
-\mathcal{E}^{\;\; \alpha\beta}_{\mu\nu} \tilde{h}_{\alpha\beta} + m^2\left(\tilde{h}_{\mu\nu}-\eta_{\mu\nu}\tilde{h}\right)  - {1 \over 2}m^2 \eta_{\mu\nu}\chi \,  
= -\Lambda\eta_{\mu\nu}\,,
\label{linhchi}
\ee
which, for the $\tilde{h}_{\mu\nu}$ solution given by~(\ref{osc}), fixes the solution for $\chi$:
\begin{equation}
\label{oscchi}
\chi \, = \,   {4\Lambda \over m^2} (\cos(mt) \, - \, 1).   
\end{equation} 
Note that this is consistent with~(\ref{tilderep}), which requires that $\chi$ be given by
\begin{equation}
\label{chiproj}
\chi \, = \,-2 \, \Pi_{\alpha\beta}\tilde{h}^{\alpha\beta}\,. 
\end{equation}
Meanwhile, the gauge-invariant metric perturbation is both transverse and has constant trace, as demanded by the Bianchi identity. 

\section{The Role of Non-linearities} 

Next we turn on non-linear interactions for $h_{\mu\nu}$. As mentioned earlier, a crucial result of self-interactions is the strong coupling phenomenon, which 
gives rise to a new scale --- the $r_\star$-radius.  In the proximity of  gravitating sources, 
the contribution from the extra helicities to the gauge-invariant metric $h_{\mu\nu}$ is suppressed, and the latter behaves almost as Einstein.  
This phenomenon takes place both in space and in time. We wish to highlight 
its nature by studying a simple model that reduces the influence of non-linearities to its bare essentials.  

Our toy non-linear model of gravity consists of restricting to exact invariance under abelian gauge transformations given in~(\ref{gauge}).
In other words, instead of non-linearly completing this gauge symmetry to full non-abelian diffeomorphism invariance, which would yield Einstein gravity,
here we study abelian non-linear gravity, for which~(\ref{gauge}) is an exact symmetry. Central to this construction is the realization that
$h_{\mu\nu}$ is itself gauge-invariant and therefore can appear in the action without derivatives. This also explains how $\Lambda$ gets
degravitated in this theory, for $h_{\mu\nu}$ is now a physical observable and hence cannot grow unbounded.

As in the spin-1 case, introducing non-linear terms does not spoil the degravitation effect already achieved in linear massive gravity. For example, consider adding
\begin{equation}
\label{term1}
(h_{\mu\nu}h^{\mu\nu}-h^2)^2\,. 
\end{equation}
In the case of a single massive graviton such a term gives rise to ghost-like instabilities for some background geometries, as was first pointed out by
Boulware and Deser~\cite{BD} and generalized in~\cite{gregoryandrei}.
In the St\"uckelberg  formalism the origin of this instability is crystal clear. Indeed, ignoring the 
spin-1 component, the kinetic terms of  $\tilde{h}_{\mu\nu}$ and $\chi$ are diagonalized 
by ~(\ref{hdecomp}).  Substituting this form into the Pauli-Fierz mass term,  it is easily seen that all higher derivative terms for $\chi$ 
cancel out  after partial integration.  This is of course why the Pauli-Fierz mass term is ghost-free.  

The nonlinear term in~(\ref{term1}) also involves higher powers of derivative interactions:
\begin{equation}
\label{hder}
 \left(m^{-4} \left(\partial_\alpha\partial_\beta\chi\right)^2 \, - \, m^{-4} \left(\Box\chi\right)^2\, 
+ \, ...\right)^2\,. 
\end{equation}
Unlike the Pauli-Fierz mass term, however, in this case the derivative self-interactions cannot be eliminated through integration
by parts. Consequently, although linear metric perturbations around flat space are ghost-free, this is not 
in general guaranteed for non-trivial backgrounds.  For the time being, we wish to disentangle 
the issue of non-linear ghost from degravitation, although, as we shall see, 
for massive gravitons some intrinsic relation between them can be established.  

Note that the story is entirely different for a resonance graviton. With $m^2(\square) \equiv \sqrt{\square}/L$, for instance,~(\ref{hder}) becomes
\begin{equation}
\label{hder2}
L^4  \left( \left({\partial_\alpha\partial_\beta \over \sqrt{\square}}\chi\right)^2 - \,  \left(\sqrt{\Box}\chi\right)^2\, 
+ \, ...\right)^2,  
\end{equation}
in which case the number of derivatives in each vertex is less than or equal to the number of fields. Nevertheless, one has to be very careful in generalizing such conclusions to 
  a full generally covariant theory, in which the number of terms is infinite. 
 
Equation~(\ref{hder}) illustrates the essence of the strong coupling phenomenon~\cite{ddgv},  since this term is singular in $m \rightarrow 0$ limit.  We now demonstrate how this strong coupling 
  leads to the $r_\star$-effect in our non-linear abelian gravity theory.

\section{Decoupling of St\"uckelberg and  $r_\star$-Effect  in Non-Linear Abelian Gravity}
\label{rstarabelian}

To shed light on the strong coupling of the helicity-0 mode and the consequent $r_{\star}$-phenomenon inherent 
to all the theories of interest, we study a toy model obtained by adding a quartic interaction~(\ref{term1}) to massive gravity.
We will see that this theory shares some of the key properties of massive gravity.

The resulting equation of motion is thus given by
\be
\label{eqnon}
\lf\mathcal{E}h\rt_{\mu\nu} - m^2\lf h_{\mu\nu} -\eta_{\mu\nu}h\rt -\frac{\lambda m^2}{6}\lf h_{\mu\nu} -\eta_{\mu\nu}h\rt \lf h_{\mu\nu}h^{\mu\nu}-h^2\rt = -T_{\mu\nu} \,.
\ee
As we did in the linear case, we decompose the metric fluctuation into helicity-2 ($\tilde{h}_{\mu\nu}$) and helicity-0 ($\chi$) modes, as given by~(\ref{hdecomp}).
While the last term in~(\ref{hdecomp}) gives a vanishing contribution in the exchange amplitude between conserved sources, nevertheless
it gives the dominant contribution in the self-interaction term as $m\rightarrow 0$, indicating that $\chi$ is strongly coupled in this limit. Thus we henceforth focus on these derivative interactions. More precisely, the decoupling limit of interest corresponds to $m\rightarrow 0$ while keeping $\lambda/m^6$ fixed, in which case~(\ref{eqnon}) simplifies to
\bea
\nonumber
& & \lf\mathcal{E}\tilde{h}\rt_{\mu\nu} - m^2\lf \tilde{h}_{\mu\nu} -\eta_{\mu\nu}\tilde{h}\rt  + \frac{1}{2}m^2\eta_{\mu\nu}\chi \\
& & \qquad -  \frac{\lambda}{6\cdot 27 m^4} \left(\partial_\mu\partial_\nu\chi -\eta_{\mu\nu}\Box\chi\right)\left(\left(\partial_\alpha\partial_\beta\chi\right)^2 - \left(\Box\chi\right)^2\right) = -T_{\mu\nu}\,.
\label{tildeeom}
\eea

Although $\chi$ is strongly coupled, we wish to show that its classical solution for a static source gives a negligible contribution to the force on a test particle at distances $r\ll r_\star$. In other words, since $\chi$ couples to the trace of the stress tensor, we will argue that the fifth force it mediates is small compared to the Newtonian force, which is dominated by $\tilde{h}$. Thus, from the outset let us assume that $\chi\ll \tilde{h}$ in~(\ref{tildeeom}); we will verify {\it a posteriori} that the consistency of this approximation. The equation of motion for $\chi$ can be obtained by taking the divergence of~(\ref{tildeeom}) and using the Bianchi identity:
\be
\partial_\nu\tilde{h} =  \frac{\lambda}{81 m^6}\partial^\mu\left\{ \left(\partial_\mu\partial_\nu\chi -\eta_{\mu\nu}\Box\chi\right)\left(\left(\partial_\alpha\partial_\beta\chi\right)^2 - \left(\Box\chi\right)^2\right)\right\}\,,
\label{divchi}
\ee
where we have assumed $\chi\ll \tilde{h}$, as mentioned earlier. 

To solve~(\ref{tildeeom}) for a static source, $T_{00} = 8\pi M\delta^3(r)$, $T_{0\mu} = 0$, a convenient ansatz is then
\bea
\nonumber
\tilde{h}_{00} &=& -\frac{M}{r}\left\{1 + {\cal O} \left(\left(\frac{r}{r_\star}\right)^p\right) \right\}\\
\nonumber
\tilde{h}_{ij} &=&  -\frac{M}{r}\delta_{ij} \left\{1 + {\cal O} \left(\left(\frac{r}{r_\star}\right)^q\right) \right\} \\
\chi & =& \frac{M}{r} {\cal O} \left(\left(\frac{r}{r_\star}\right)^s\right) \,,
\label{ansatz}
\eea
where  $p,q,s$ are unknown positive powers, and $r_\star$ is to be determined below. Anticipating the decoupling of $\chi$ for $r\ll r_\star$, this ansatz reduces to the standard Newtonian solution for a static source in this limit. Thus, to leading order we have $\tilde{h} \approx -2M/r$, and therefore $\partial_i \tilde{h} = 2M x_i/r^3$. Substituting into~(\ref{divchi}), by inspection we deduce that 
\be
\chi \sim r^{5/3}\,,
\ee
where the constant of proportionality is fixed by explicit substitution. The final result is
\be
\chi = -\frac{27}{5} \left(\frac{m^6M}{28\lambda}\right)^{1/3}r^{5/3}\,.
\label{localchi}
\ee

It remains to show that $\chi\ll \tilde{h}$ at small distances, as assumed throughout. This not only justifies the approximation made earlier, as well as the ansatz~(\ref{ansatz}), but more importantly immediately implies that the force on a test particle is dominated by the Newtonian contribution. We find
\be
\label{ratio}
\frac{\chi}{\tilde{h}}\sim r^{8/3}\left(\frac{m^6}{M^2\lambda}\right)\equiv \left(\frac{r}{r_\star}\right)^{8/3}\,,
\ee
where
\be
r_\star \equiv \left(\frac{M^2\lambda}{m^6}\right)^{1/8} \,.
\ee
Thus, as claimed, the $\chi$ contribution is suppressed at distances $r\ll r_\star$, and the solution well approximates Einstein gravity.

As an alternative derivation, we can instead solve the decoupled equations of motions 
 for $\tilde{h}_{\mu\nu}$ and $\chi$. The equation of motion for the latter is obtained by substituting~(\ref{hdecomp}) in the action 
 and  taking variation with respect to $\chi$. This is equivalent to acting on~ (\ref{eqnon}) with the operator $\eta_{\mu\nu}/6 \,  + \, \partial_\mu\partial_\nu/3m^2$.
  In the decoupling limit, we thus obtain the decoupled equations,
\be
\label{hdec}
\lf\mathcal{E}\tilde{h}\rt_{\mu\nu} \, = -T_{\mu\nu} \,,
\ee
and 
\be
\label{chidec}
\square \chi \, 
  -  \frac{\lambda}{81 m^6} \left(\partial_\mu\partial_\nu\chi -\eta_{\mu\nu}\Box\chi\right)\partial^{\mu}\partial^{\nu} \left(\left(\partial_\alpha\partial_\beta\chi\right)^2 - \left(\Box\chi\right)^2\right) = -T\,,
 \ee
which are equivalent, again in the decoupling limit, to~(\ref{tildeeom}) and~(\ref{divchi}). 
In particular, at short distances we can neglect the first term on the left hand side of~(\ref{chidec}). Then, assuming that $\chi \propto r^n$, we find
that the non-linear term behaves as  $8n^2 (1-2n)(n-2) (5-3n)\, r^{3n-8}$, which matches the 
delta function source for $n = 5/3$. Meanwhile, from~(\ref{hdec}) we deduced that $h_{00}\sim 1/r$, 
in agreement with~(\ref{ratio}).

  Although the relative suppression of $\chi$ for $r \ll r_{\star}$ is generic,  we should stress that matching with the $1/r$ solution at large distances is not guaranteed in all theories with strong coupling~\footnote{The absence of the interpolating static solutions in massive gravity, is also supported by the analysis of~\cite{kogan}.}. This is because the high-derivative nature of non-linear interactions can trigger ghost-like instabilities in the presence of a source~\cite{ghostnicolis}. Indeed, the suppression of 
 $\chi$ near a source can be interpreted as arising from the exchange of a ghost at short distances~\cite{ghostcedric}. 
 We shall encounter the time-like version of this phenomenon when discussing the phase transitions, in which case an exponentially growing mode gets excited.

 \section{Degravitation in Non-Linear Abelian Gravity}
 
We now turn to the study of degravitation in the framework of non-linear abelian gravity. First we focus on a pure cosmological term and show that, for finite $m$ and $\lambda$, any constant
vacuum energy is effectively degravitated and yields a flat metric. Then we discuss phase transitions involving a time-dependent yet conserved energy-momentum tensor. This
yields a non-trivial metric, describing the transition from one Minkowski vacuum to the other, with subsequent oscillations about the latter. 

Next we study degravitation in the decoupling limit. Even for $m^{-1}\rightarrow\infty$, there is still filtering. Nevertheless, we refer to degravitation in this limit as {\sl neutralization}. When sourced by a cosmological term, $\tilde{h}_{\mu\nu}$ and $\chi$ both acquire non-trivial solutions. But their contributions to the full gauge-invariant metric precisely cancel out, such as to yield a flat $h_{\mu\nu}$. Furthermore, the analysis of phase transitions reveals a $t_\star$-effect for $\chi$, which is the time-like analogue of the $r_\star$-phenomenon discussed in the previous section. More precisely we will see that the strong coupling of the helicity-0 mode results in suppressed response shortly after the phase transition. We find an intriguing relation between the $t_\star$ delay and the excitation of the BD ghost, which is present in non-linear abelian gravity around non-trivial backgrounds.

We then study degravitation of the cosmological constant without taking the decoupling limit, that is, for finite $m^{-1}$. In the limit $t_\star \gg m^{-1}$, the dominant degravitation mechanism is through finite-$m$ filtering, as opposed to neutralization. This analysis allows us to check some of the claims made in the introduction. For instance, in this filtering regime $\chi$ is nearly decoupled and its role is limited to maintaining the Bianchi identity. As in linear Fiertz-Pauli, we derive oscillating solutions around the new Minkowski vacuum. 

\subsection{Response to Phase Transitions}

Thus let us start with~(\ref{eqnon}), focusing on the case of pure vacuum energy: $T_{\mu\nu}\, = \, -\Lambda\eta_{\mu\nu} $. Evidently, this is solved by 
\be
h_{\mu\nu} \, = \, f_0 \eta_{\mu\nu} \,, 
\ee
where the constant $f_0$ satisfies
\begin{equation}
\label{fzero}
f_0  \,  - \, 2 \lambda \, f_0^3 \, = \, {\Lambda \over 3m^2}\,.  
\end{equation}
Hence, flat Minkowski space is a solution for arbitrary $\Lambda$. As advocated, non-linear effects do not spoil degravitation, exactly as in the spin-1 case. 

It is important to note, however, that around this shifted
Minkowski vacuum the mass term of metric fluctuations receives
corrections from the non-linear term which are not of the PF form.
This signals that on the new background the helicity-0
graviton propagates the ghost.  For consistency of the analysis,
the corresponding pole must lie at momenta above the cutoff.   

To see what this entails for the parameters of the theory, let us assume for concreteness that $\lambda$ is small such that the linear term dominates in~(\ref{fzero}):
\be
f_0\sim {\Lambda \over m_{\rm Pl}^2m^2}\,,
\label{f0lin}
\ee
where we have restored $m_{\rm Pl}^{-2}\equiv 16\pi G$. (As usual $\Lambda$ has dimension 4, so $f_0$ is dimensionless, as it should.) This shift
the mass by order $\delta m^2 \, \sim \, \lambda \; \Lambda^2/m_{\rm Pl}^2m^2$, leading to a ghost pole at
\be
m_{\rm g}^2 \, \sim \, {m^4 \over \delta m^2} \, =\, {\Lambda_{\rm strong}^8 m_{\rm Pl}^2 \over \Lambda^2}\,,
\label{pole}
\ee
where 
\be
\Lambda_{\rm strong}
\, \equiv \, \left(\frac{m_{\rm Pl}^2m^6}{\lambda}\right)^{1/8}
\ee
is the strong coupling scale for $\chi$. Requiring that the ghost pole lies above the $\Lambda_{\rm strong}$ cut-off implies
\be
\Lambda \ll \Lambda_{\rm strong}^3 m_{\rm Pl}\,.
\label{bound3}
\ee
Meanwhile, the linear approximation~(\ref{f0lin}) is valid provided $\lambda f_0^2\ll 1$, which in turn implies $\Lambda \ll \Lambda_{\rm strong}^4 m_{\rm Pl}/m$. But
this immediately follows from~(\ref{bound3}) for $\Lambda_{\rm strong}\gg m$, which is the case of interest. A similar analysis can be done for the opposite regime in which
the non-linear term in~(\ref{fzero}) dominates, yielding analogous bounds on $\Lambda$. We will come back to the ghost when we encounter the $t_{\star}$-effect for phase transitions, in which, as we will see, the ghost plays a non-trivial role.

We now wish to investigate how the metric responds to phase transitions 
described by a homogeneous and isotropic source.  We thus take $T_{\mu\nu}$ to be time-dependent only, as well and diagonal  
\begin{equation}
\label{sourcetimedep}
T_{\mu\nu}  \, = \, {\rm diag} (\rho, P(t), P(t), P(t)) \,.
\end{equation}
Conservation of the source, $\partial^{\mu} T_{\mu\nu} \, = \, 0$, of course forces $\rho$ to be constant. Thus our phase transitions can only involve a change in pressure. 

As ansatz, we look for solutions of the form 
\be
h_{00} \, = \, g(t), ~~h_{0i} \, = \, 0, ~~h_{ij} \, = \,  \delta_{ij} \, f(t)\,.
\label{gf}
\ee
Taking the divergence of~(\ref{eqnon}) and using Bianchi identity, we get the following constraint: 
\begin{equation}
\label{consf}
\partial^{\mu} \left \{ (h_{\mu\nu} -\eta_{\mu\nu}h) \, \left[1 + \lambda f(g -  f) \right ] \right\} \, = \, 0\,,
\end{equation}
which implies 
\begin{equation}
\label{consf1}
f  \,\left[1 + \lambda f(g -  f) \right ]  \, = \, {\rm const}\,. 
\end{equation}
From the $00$-equation of~(\ref{eqnon}) , this constant is fixed to be $-\rho/3m^2$. Thus, we can solve for $g(t)$:
\begin{equation}
\label{gsol}
g(t) \, = \,-\, {\rho \over 3m^2 \lambda f^3(t) }\, -\, {1 \over 3\lambda f(t)}  \, + \, f(t)\,.
\end{equation} 

\begin{figure}[ht]
\includegraphics[width=100mm]{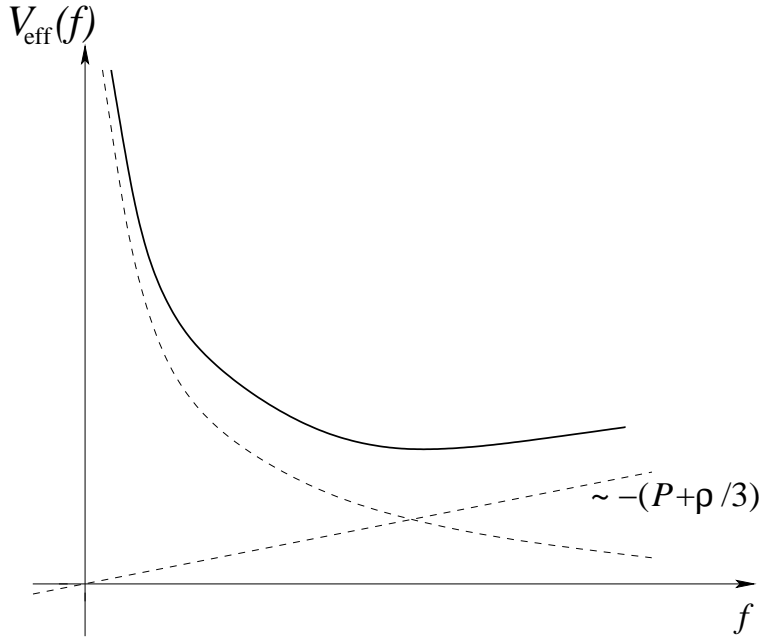}
\centering
\caption{The effective potential for the metric function $f(t)$, as given by~(\ref{fpot}).}
\label{veff}
\end{figure}

Meanwhile, the $ii$-equation reduces to
\begin{equation}
\label{ijeq}
2\ddot{f}(t) \,-\, \left ( {\rho \over 9m^2 \lambda f^3(t) } \, + \, {1 \over 3\lambda f(t)}  
+ {1 \over 3} \right ) \rho \, = \, P(t) \,,
\end{equation}
which describes the evolution of a time-dependent scalar field $f(t)$ in an effective potential:
\begin{equation}
\label{fpot}
V_{\rm eff}(f) \, =  \,  {\rho^2  \over 36 m^2 \lambda f^2(t) }\, - \, {1 \over 6\lambda} {\rm ln} ( f(t))\,  
- \, {1 \over 2}  f(t) \, \left(\frac{\rho}{3} \, + \, P(t)\right) \;.
\end{equation}
See Fig.~\ref{veff}. In particular, $f$ is effectively sourced by $ - \, (\rho/3 \, + \, P(t))$. Since $\lambda >0$, then for any constant
$P$ satisfying $P < -\rho/3$, the effective potential displays a stable minimum at $f \, = \, f_0$, where $f_0$ satisfies
\begin{equation}
\label{fzero2}
 {\rho \over 9m^2 \lambda  f_0^3} \, + \, {1 \over 3\lambda f_0} 
=  {\rho \over 3} \, - \, P\,.
\end{equation}
Whenever the energy density is dominated by vacuum energy, we have $P\simeq -\rho$,
which therefore generates a stable point for $f$. (Of course for $P=-\rho = -\Lambda$,~(\ref{fzero2})
reduces to~(\ref{fzero}), as it should.)

Consider now a phase transition from one vacuum-energy dominated state to another, during which 
 the pressure changes sharply by a small amount. For concreteness let us take $P(t) \, = \, -\rho  \, + \, \Delta\theta(t)$, where  $\Delta \ll \rho$.
 The transition then amounts to a small instantaneous shift in the position of the  minimum. 
 If the system is initially at rest in the original minimum, then the transition at  $t = 0$ will be followed by oscillations around the new minimum, which in the presence of friction ({\it e.g.}, due to particle creation) will damp out.
 
 \subsection{Decoupling Limit, Neutralization and $t_\star$-Effect} 
 
Next we study degravitation in the decoupling limit: $m\rightarrow 0$, $\lambda\rightarrow 0$, keeping $\lambda/m^6$ fixed. In this limit the equations for $\tilde{h}$ and $\chi$ decouple and reduce to~(\ref{hdec})  and~(\ref{chidec}), respectively, which we include here for convenience:
\bea
\label{dec2}
\nonumber
& & \lf\mathcal{E}\tilde{h}\rt_{\mu\nu} \, = -T_{\mu\nu} \,; \\ 
& & \square \chi \, 
  -  \frac{\lambda}{81 m^6} \left(\partial_\mu\partial_\nu\chi -\eta_{\mu\nu}\Box\chi\right)\partial^{\mu}\partial^{\nu} \left(\left(\partial_\alpha\partial_\beta\chi\right)^2 - \left(\Box\chi\right)^2\right) = -T\,.
 \eea

Starting once again with a pure cosmological term, $T_{\mu\nu}\, = \, -\Lambda \eta_{\mu\nu}$, clearly a solution for $\tilde{h}$ is just linearized de Sitter
\be
\tilde{h}_{\mu\nu} \, = \, -{\Lambda \over 12} \eta_{\mu\nu} x_{\alpha}x^{\alpha}\,.
\label{hcc1}
\ee
This is not surprising since we have effectively sent the filter time-scale $m^{-1}$ to infinity in this decoupling limit, thereby neutralizing any filtering effect on $\tilde{h}_{\mu\nu}$.
There is, nevertheless, degravitation as far as the full metric $h_{\mu\nu}$ is concerned. This is degravitation not because of filtering, but instead due to the contribution from the helicity-0 mode.
Indeed, a solution for $\chi$ is
\be
\chi \, = \, {\Lambda \over 2}  x_{\alpha}x^{\alpha}\,.  
\label{chicc1}
\end{equation}
When substituted in the full metric decomposition, $h_{\mu\nu} = \tilde{h}_{\mu\nu} + \eta_{\mu\nu}\chi/6 +\; {\rm gauge}\; {\rm terms}$, we see that~(\ref{hcc1}) and~(\ref{chicc1}) precisely cancel, yielding a
flat space solution for $h_{\mu\nu}$.

Thus the decoupling limit still leads to the degravitation mechanism, relying on a cancellation between helicity-2 and helicity-0 contributions to the physical metric. This can be understood
as a limit of filtering, which we henceforth refer to as {\sl neutralization}. For finite $m$, both neutralization and standard filtering contribute to degravitation. Which mechanism dominates depends
on the time scale for each channel. The time scale for filtering is of course $m^{-1}$. Meanwhile, as we will see below in the study of phase transitions, the relevant time scale for neutralization
is $t_\star$, a time-like analogue of $r_\star$ for spatially-localized sources. Thus, depending on whether $t_\star\gg m^{-1}$ or $t_\star\ll m^{-1}$, degravitation will predominantly take place through standard filtering or neutralization, respectively. 

Let us therefore turn to the study of phase transitions. We will find that, shortly after the phase transition, the strong coupling of $\chi$ suppresses its response compared
to that of $\tilde{h}$. Since phase transitions are described by sources localized in time, this strong coupling suppression is a time-like analogue of the $r_\star$-effect
near spatially-localized sources. 

Working once again in the decoupling limit, consider~(\ref{dec2}) with the following source term
\begin{equation}
\label{tforcc}
T_{\mu\nu}\, = \, -\Lambda \eta_{\mu\nu}  \, + \, \delta_{\mu}^i\delta_{\nu}^j\delta_{ij} \, \epsilon(t)\,,
\end{equation}
corresponding to a cosmological term plus a small change in pressure described by $\epsilon(t)$. The solutions for the helicity-2 and -0 states are given respectively by
\begin{equation}
\label{hsolcc}
\tilde{h}_{\mu\nu} \, = \, -{\Lambda \over 12} \eta_{\mu\nu} x_{\alpha}x^{\alpha} \, + \, \delta_{\mu}^i\delta_{\nu}^j \delta_{ij} q(t)\,,
\end{equation}
and 
\begin{equation}
\label{chisolcc}
\chi \, = \, { \Lambda \over 2} x_{\alpha}x^{\alpha}\, + \, r(t)\,,
\end{equation}
where the leading terms proportional to $\Lambda$ take care of the cosmological term as in~(\ref{hcc1}) and~(\ref{chicc1}). The functions $q(t)$ and $r(t)$ are small corrections
encoding the metric response to the phase transition. They satisfy, respectively,
\begin{equation}
\label{qoft}
\frac{d^2q}{dt^2} \, = \, \epsilon(t) \;;\qquad  \frac{d^2r}{dt^2}\, + \, {2 \lambda \Lambda^2 \over 9m^6} \,  \frac{d^4r}{dt^4} \, = \, 3 \epsilon(t)  \; .
\end{equation}

In order to highlight the $t_\star$-effect, we focus on transitions that are sharply 
localized in time, for instance $\epsilon(t) \, = \, \Delta \theta(t)$, where $\Delta$ is constant.  
It immediately follows that 
\be
q(t) \, = \, {1 \over 2} \Delta \theta(t) \, t^2\,.
\ee
Thus $\tilde{h}$ grows as $t^2$ after the phase transition. Meanwhile, an exact solution for $r$ is
\begin{equation}
\label{finstab}
r(t) \, = \,  3 \Delta \theta(t) \, \left [ {2\lambda \Lambda^2 \over  9m^6}\, \left (\cos 
\left( t\sqrt{{9m^6 \over 2\lambda\Lambda^2}} \right ) \, - 1 \right ) \, + \, {t^2 \over 2}\right ]   \;,
\end{equation}
which, for 
\begin{equation}
\label{tstar}
t \, \ll \, t_\star \, \equiv \, \sqrt{{\lambda \Lambda^2 \over m^6}} \;,
\end{equation} 
behaves as
\begin{equation}
\label{flesst}
r(t) \, \simeq\, { 9 \over 16} \, {\Delta m^6 \over \lambda \Lambda^2}\, \theta(t) \, t^4\,.
\end{equation} 

Therefore, for $t \ll t_\star$ the contribution from $\chi$ to the full metric $h_{\mu\nu}$ is suppressed. 
Indeed, substituting the solutions for $\tilde{h}_{\mu\nu}$  and  $\chi$ in~(\ref{hdecomp}), we obtain
\begin{equation}
\label{hlesst}
h_{\mu\nu} \, = \, \delta_{\mu}^i\delta_{\nu}^j\delta_{ij} \left(\frac{q(t)}{2} \, + \,\frac{r(t)}{6}\right) \, + \, {\rm pure~gauge~terms}.
\end{equation}
Since
\begin{equation}
\label{foverq}
{r(t) \over q(t)} \, \sim {t^2 \over t_\star^2} \, \ll \, 1,
\end{equation}
this implies that on short time scales, the metric responds to the phase transition almost as in (linear) Einstein gravity,  
with small corrections coming from $\chi$.  This $t_\star$-phenomenon is the time-like counterpart of the $r_\star$-effect for localized
sources discussed earlier.

The behaviour of the solution for $t > t_\star$ depends on the sign of 
$\lambda$. For positive $\lambda$, the first term in the solution is 
oscillatory, and in this case the contributions from $r(t)$ and $q(t)$ in the metric cancel up to an 
oscillatory part. 

For negative $\lambda$, there is a tachionic growth. 
Thus the solution~(\ref{finstab}) for $\chi$ never recovers the naive linear regime with 
$r(t) \propto  t^2$, which would have obtained had we ignored non-linearities. Instead, at late times ($t \gg t_\star$), the 
contribution from  $\chi$ not only `catches up' with that of $\tilde{h}_{\mu\nu}$, but overshoots and continues to grow exponentially. 
The origin of this instability lies in the high-derivative nature of the non-linear term and 
traces back to the Boulware-Deser instability.  Indeed, in the absence of the transition ($\epsilon(t) \, = \, 0$), small perturbations around the flat background,
\begin{equation}
\label{smallpert}
\chi \, = \, { \Lambda \over 2} x_{\alpha}x^{\alpha}\, + \, \delta\chi (x)\,,  
\end{equation}
satisfy the equation of motion
\begin{equation}
\label{smalleq}
\square \delta\chi \, - \, {2  \Lambda^2 \over 9 m_{\rm Pl}^2 \Lambda_{\rm strong}^8 } \,  \square^2\, \delta\chi \, = \, 0\,.
\end{equation}
The higher-derivative nature of this equation implies that $\delta\chi$ propagates a massless scalar, as well as a ghost of mass
\begin{equation}
\label{pole1}
m_{\rm g}^2 \, =\, {9 m_{\rm Pl}^2\Lambda_{\rm strong}^8  \over
2\; \Lambda^2}\,,
\end{equation}
in agreement with~(\ref{pole}).  
Comparison with~(\ref{qoft}) reveals that both the oscillation (for $\lambda > 0$) and 
exponential growth (for $\lambda < 0$) we have encountered is caused by 
the fact that the phase transition sources both ordinary scalar and a 
ghost. The recovery of the approximately
Einsteinian metric in the vicinity of the transition, $t \ll t_{\star}$, can thus be understood as 
the compensation of the normal scalar by ghost exchange. This is very similar to the story for spatially localized sources suggested by
Deffayet and Rombouts~\cite{ghostcedric}.  Once the instability sets in, however, it continues to manifest itself at  much later times than $t_{\star}$, again in analogy with the space-like
story of~\cite{ghostnicolis}.  

Thus the lesson is that in massive gravity the $t_{\star}$-phenomenon and BD ghost-like mode are 
intrinsically related.  In Sec.~\ref{nonabelian} we will see that a similar effect takes place in the decoupling limit of non-abelian massive gravity.

\subsection{Degravitation for Finite Filter Scale}
\label{degravfilter}

In the previous subsection we focused on degravitation in the decoupling limit of non-linear abelian gravity. In this limit, filtering becomes irrelevant, and instead degravitation
takes place through neutralization --- the cancellation of helicity-2 and helicity-0 components to yield a flat physical metric. More generally, neutralization is the dominant
degravitating mechanism whenever $t_\star \ll m^{-1}$. We now wish to study the opposite regime, $t_\star \ll m^{-1}$, in which filtering is efficient. 

In essence we seek for non-linear generalizations to the solutions of Sec.~\ref{spin2} derived in linear Pauli-Fiertz gravity. Recall that these solutions approximate de Sitter at early times and then oscillate with frequency $m^{-1}$ around the new flat space vacuum. Here we wish to understand how the strong coupling of the longitudinal mode affects the story. 

More precisely, we treat $m$ as a small expansion parameter and look for the leading filtering correction to de Sitter geometry. Let us therefore expand the helicity-2 and helicity-0 components as
\be
\tilde{h}_{\mu\nu} = \tilde{h}_{\mu\nu}^{(0)} + \tilde{h}_{\mu\nu}^{(1)} +\ldots \;;\qquad \chi = \chi^{(0)}+\ldots\,,
\ee
where the superscript indicates the perturbative order in $m^2$. Substituting into~(\ref{eqnon}) and grouping terms of the same order, we obtain
\be
\left\{\lf\mathcal{E}\tilde{h}^{(0)}\rt_{\mu\nu} - \Lambda\eta_{\mu\nu}\right\} +m^2\left\{\frac{1}{m^2}\lf\mathcal{E}\tilde{h}^{(1)}\rt_{\mu\nu} - \lf \tilde{h}_{\mu\nu}^{(0)}-\tilde{h}^{(0)}\eta_{\mu\nu}\rt +\frac{\chi}{2}\eta_{\mu\nu}- K_{\mu\nu}\right\} = 0 \,,
\label{expandeom}
\ee
where $K_{\mu\nu}$ includes all non-linear terms. Although $m$ is assumed small, let us also take $\lambda$ to be small such that $\chi$ is strongly coupled. In this near decoupling limit, we can keep only the non-linearities in $\chi$, while treating $\tilde{h}$ linearly, in which cases
\be
K_{\mu\nu} = \frac{\lambda}{192\; m^6}\lf\partial_\mu\partial_\nu\chi-\eta_{\mu\nu}\Box\chi\rt\lf\lf\partial_\alpha\partial_\beta\chi\rt^2-\lf\Box\chi\rt^2\rt\,.
\label{K1}
\ee

The zeroth-order equation, $\lf\mathcal{E}\tilde{h}^{(0)}\rt_{\mu\nu} = \Lambda\eta_{\mu\nu}$, has for its solution linearized de Sitter space:
\be
\tilde{h}_{\mu\nu}^{(0)} = -\frac{\Lambda}{12} \eta_{\mu\nu}x^2\,,
\label{h0}
\ee
where $x^2\equiv x_\alpha x^\alpha $. Next we can solve for $\chi^{(0)}$ by taking the divergence of~(\ref{expandeom}) and using the Bianchi identity. In the near decoupling limit of interest, we anticipate that it is consistent to neglect the $\chi$ mass term in this equation, which amounts to $\chi^{(0)}\ll \tilde{h}^{(0)}$. At the end of the day we will come back and check the validity of this approximation. The divergence of~(\ref{expandeom}) thus yields
\be
\partial^\mu K_{\mu\nu} \approx -\partial^\mu\lf \tilde{h}_{\mu\nu}^{(0)}-\tilde{h}^{(0)}\eta_{\mu\nu}\rt = -\frac{\Lambda}{2}x_\nu\,,
\ee
whose general solution for $K$ is of the form
\be
K_{\mu\nu} = \frac{1}{2}\Lambda\gamma\lf -x_\mu x_\nu +\lf 1-\frac{1}{2\gamma}\rt x^2\eta_{\mu\nu}\rt\,,
\label{K2}
\ee
where $\gamma$ is an arbitrary constant to be fixed shortly. 

In order to reproduce the $x^2$-dependence of $K$, we infer from~(\ref{K1}) that $\chi^{(0)}\sim \lf x^2 \rt^{4/3}$. For instance, 
\be
\chi^{(0)} = -\frac{3}{4}\lf\frac{\Lambda m^6}{8\lambda}\rt^{1/3}\lf x^2 \rt^{4/3}
\label{chi0}
\ee
generates a $K_{\mu\nu}$ of the form~(\ref{K2}) with $\gamma = -1/9$. This solution closely resembles~(\ref{localchi}) for the localized source. Indeed in both cases $\chi$ grows as some rational power of the coordinate distance, and does so more slowly than its helicity-2 cousin. Thus, already we can foresee the decoupling of the longitudinal mode at short distances. 

Next we substitute the expression for $K_{\mu\nu}$ into~(\ref{expandeom}) to obtain an equation for the leading correction in $\tilde{h}$:
\be
\lf\mathcal{E}\tilde{h}^{(1)}\rt_{\mu\nu} = m^2\lf \tilde{h}_{\mu\nu}^{(0)}-\tilde{h}^{(0)}\eta_{\mu\nu} +K_{\mu\nu} \rt  = \frac{m^2\Lambda}{18}\lf x_\mu x_\nu - x^2 \eta_{\mu\nu}\rt\,.
\ee
Evidently, the backreaction of $\chi$, encoded in $K$, amounts to a correction to the background mass term. It follows that
\be
\tilde{h}^{(1)}\sim \cO\lf m^2\Lambda x^4\rt\,,
\ee
which, combined with~(\ref{h0}), reveals a filtering solution analogous to~(\ref{osc}):
\be
\tilde{h} =  -\frac{\Lambda}{3}x^2+ \cO\lf m^2\Lambda x^4\rt\approx \frac{\Lambda}{2m^2}\lf 1-\cos m\sqrt{x^2}\rt\,.
\ee
Thus, as expected, the non-linearities do not spoil the oscillating solutions derived in linear Pauli-Fiertz.

It remains to check the consistency of the decoupling approximations. First it is easily seen that $\chi$ is negligible at short distances compared to the helicity-2 component:
\be
\frac{\chi}{\tilde{h}^{(0)}} \sim  \lf\frac{m^6}{\lambda\Lambda^2}\rt^{1/3}\lf x^2 \rt^{1/3} \equiv \lf\frac{x^2}{t_\star^2}\rt^{1/3}\,,
\ee
where $t_\star$ is defined in~(\ref{tstar}). (Here, however, the latter is probably best understood as some $x_\star$.) What is more interesting is that $\chi$ can even be suppressed compared
to $\tilde{h}^{(1)}$:
\be
\frac{\chi}{\tilde{h}^{(1)}} \sim \frac{1}{\lf m^2x^2\rt^{2/3}}\frac{1}{\lf m^2 t_\star^2\rt^{1/3}}\,.
\ee
Indeed, although $m^2x^2\ll 1$ by assumption, this ratio can be made small provided that
\be
t_\star \gg m^{-1}\,.
\ee
But this is precisely the regime of interest in which filtering is the dominant degravitation channel over neutralization.
In any realistic cosmology $t_\star$ will be of the order of the Hubble radius, $H^{-1}$. This confirms our earlier claims that the effective filtering equation~(\ref{central}) is a valid approximate equation of motion for the {\it full} metric $h_{\mu\nu}$ provided that $H^{-1}\ll m^{-1}$. In this regime $\chi$ is strongly coupled and as a result decouples from conserved sources. Its role is limited to enforcing the Bianchi identity, which, as we have seen explicitly, merely amounts to an order one shift in the mass term.

\section{Degravitation in Non-abelian Massive Gravity}
\label{nonabelian}

 We now discuss degravitation in theories of massive gravity with full non-abelian  general covariance. The unique, consistent and fully generally-covariant  theory of a single massive graviton is of course unknown. So we shall only demonstrate the degravitation phenomenon in a certain decoupling limit, 
 which, under reasonable assumptions, is fairly insensitive to the concrete form of the completion.  

Before doing so let us reiterate that in linear PF gravity, degravitation works even in the decoupling limit, which in the linear case simply amounts to taking $m^2 \rightarrow 0$. 
This is because of the neutralization effect discussed earlier. Recall that in this limit the equations for the helicity-2 and helicity-0 components decouple --- the helicity-1 mode is of course irrelevant --- and reduce to
\be
\label{hdeclin}
\lf\mathcal{E}\tilde{h}\rt_{\mu\nu} \, = \, - \, T_{\mu\nu} \,;\qquad \square \chi \,   = \,  -T\,.
\ee
The solutions for these modes are then given by~(\ref{hcc1}) and~(\ref{chicc1}), which cancel (or neutralize) each other in the full metric.
Thus, an observer probing such a metric with an arbitrary conserved source $t_{\mu\nu}$ 
will observe flat space,
\begin{equation}
\label{observer}
\int \, d^4 x\, t^{\mu\nu}\, h_{\mu\nu} \, = \, 0\,,
\end{equation}
despite the presence of a non-zero cosmological term.

Let us now turn to the non-abelian completion of the gauge symmetry.  In this case,
as explained above,  the helicity-0  St\"uckelberg exhibits strong coupling due to interactions that are 
singular in $1/m^2$.  As shown in~\cite{ddgv,strong} the leading singularity is of order $1/m^4$ and comes from the 
trilinear vertex. Hence the interactions become strongly coupled at $\Lambda_{\rm strong} \equiv (m^4m_{\rm Pl})^{1/5}$. 
We will study this system by taking the useful decoupling limit of~\cite{strong}: 
$m \rightarrow 0,~\Lambda_{\rm strong} \, = \,$~fixed, in which only the $1/m^4$ non-linearity survives. 
We shall, however, also keep the equation for  the helicity-2 part, since the latter contributes to the full metric.  
Hence, in the decoupling limit  the equations for helicity-2 and helicity-0 components become, respectively,  
\bea
\label{hdecmass}
\nonumber
& & \lf\mathcal{E}\tilde{h}\rt_{\mu\nu} \, =  \,- {T_{\mu\nu}  \over m_{\rm Pl}}\,; \\
& & \square \chi \, 
 + {1 \over \Lambda_{\rm strong}^5}  \left ( 3 \square (\square \chi)^2  \, - \, \square (\partial_\mu\partial_\nu\chi)^2\, - \, 2 \partial^{\mu}\partial^{\nu} (\partial_{\mu}\partial_{\nu}\chi \square \chi) \right) \, =  \, -{T \over m_{\rm Pl}} \,,
\eea
where we are implicitly assuming canonical normalization for $\tilde{h}$ and $\chi$, such that they both have dimension of mass. (Correspondingly, we have also restored factors of $m_{\rm Pl}$ to make dimensions explicit.) It is immediately clear that the degravitated solutions~(\ref{hcc1}) and~(\ref{chicc1}) still hold in the presence of the non-linear term, since the latter is identically zero for $\chi$ proportional to $x_\alpha x^\alpha$. Thus, thanks to the higher-derivative structure of the non-linear term, the linearized solution just goes through. 

The non-linear term does play an important role in phase transitions, however, as it generates an instability
akin to the one uncovered in abelian gravity. To see this, consider as in~(\ref{tforcc}) adding a small transition in the pressure:
 \begin{equation}
\label{tforcc2}
T_{\mu\nu} \,  = \, -\Lambda \eta_{\mu\nu}  \, + \, \delta_{\mu}^i\delta_{\nu}^j\delta_{ij} \, \Delta \theta(t) \,.
\end{equation}
A useful ansatz for $\tilde{h}$ and $\chi$ is then
\bea
\tilde{h}_{\mu\nu} \,& = & \, -{\Lambda \over 12 m_{\rm Pl}} \eta_{\mu\nu} x_{\alpha}x^{\alpha} \, + \, \delta_{\mu}^i\delta_{\nu}^j \delta_{ij} q(t)\,;\\
\chi \, & = & \, {\Lambda \over 2m_{\rm Pl}}  x_{\alpha}x^{\alpha}\, + \, r(t)\,,  
\eea
where $q(t)$ is once again governed by the first of~(\ref{qoft}), while $r(t)$ satisfies
\begin{equation}
\label{foftnew}
\frac{d^2r}{dt^2} \, - \, {12\Lambda\over  m_{\rm Pl} \Lambda^5_{\rm strong} } \,  \frac{d^4r}{dt^4}\, = \, 3 \theta(t) \frac{\Delta}{m_{\rm Pl}} 
\end{equation}
The solution for $r(t)$ for $t \, > \, 0$ is
\begin{equation}
\label{fch}
r(t) \, = \,  {3 \Delta \over 2 m_{\rm Pl}} \, \left [ -{24\Lambda \over  m_{\rm Pl}\Lambda^5_{\rm strong} }\, \left ({\rm ch} \left( t\sqrt{{ m_{\rm Pl}\Lambda_{\rm strong}^5 \over 12\Lambda}} \right ) \, - 1 \right )\, + \, t^2 \right ]\,,   
\end{equation}
which, for 
\begin{equation}
\label{tstar1}
t \, \ll \, t_\star \, \equiv \, \sqrt{{12\Lambda \over  m_{\rm Pl}\Lambda^5_{\rm strong} }} \,,
\end{equation} 
behaves as 
\begin{equation}
\label{flesst2}
r(t) \, = \, -{ \Delta \over 96} \, {{\Lambda_{\rm strong}^5 \over \Lambda}}\, \theta(t) \, t^4  \,.
\end{equation} 

The physics is exactly as in abelian non-linear gravity. In this decoupling limit, we have derived a $t_\star$-effect: 
for $t\ll t_\star$, the response of the metric to the phase transition is approximately as in Einstein gravity. 
However, at later times the linear solution, with both $\tilde{h}$ and $\chi$ going as $x^2$, is never recovered.
Instead, $\chi$ grows unbounded, with characteristic time $t_\star$, due to an instability in $\chi$ triggered at the transition.

As in abelian gravity, the root of this instability lies in the Boulware-Deser sixth mode. 
Indeed, in the absence of the transition ($\epsilon(t) \, = \, 0$),  small 
perturbations about the flat background, $\chi \, = \, \Lambda  x^2/2m_{\rm Pl}\, + \, \delta\chi (x)$,   
satisfy
\begin{equation}
\label{smalleq2}
\square \delta\chi \, + \, {12 \Lambda  \over m_{\rm Pl}\Lambda_{\rm strong}^5} \,  \square^2\, \delta\chi \, = \, 0\,.
\end{equation}
Thus, once again  $\delta\chi$ propagates a massless scalar, as well as a ghost with mass determined
by $t_\star$:
\begin{equation}
m_{\rm g}^2 \, = \, {m_{\rm Pl}\Lambda_{\rm strong}^5 \over
12 \Lambda}\, = \, t_{\star}^{-2}\,.
\end{equation}
The phase transition therefore excites both modes, thereby triggering the instability. The recovery of an almost
Einsteinian metric for $t\ll t_\star$ can be understood as the compensation between normal scalar and ghost exchange, 
in complete analogy with the space-like story of~\cite{ghostnicolis,ghostcedric}. Notice that, although the ghost gets excited in the transition,
the exponential growth versus oscillatory behavior is determined by the sign of $\Lambda$.

\section{Degravitation and Decoupling Limit in Theories with Resonance Graviton}

 As pointed out in~\cite{gd}, the strong coupling regime discovered in~\cite{ddgv}  for massive gravity and DGP must be shared by all consistent 
 theories of large distance modified gravity with $m^2(\square) \, = \, L^{2(\alpha - 1)} \square^{\alpha}$.  As explained earlier, while $\alpha = 0$ 
 corresponds to the case of a massive graviton, 
 theories with generic $\alpha$  can be viewed as theories of a resonance graviton, whose width is set by the inverse filter scale, $1/L$.  
 In these theories, the most singular terms for large  $L$ are related to the trilinear helicity-0 vertex and 
 amount to a momentum dependence of the form (\ref{vertex}), corresponding to the strong coupling 
 scale 
 \begin{equation}
\label{strongalpha}
\Lambda_{\rm strong}\, = \, (L^{4(\alpha \, - \,1)}m_{\rm Pl})^{{1 \over 5\, - \, 4\alpha}}\,. 
\end{equation}
Due to the existence of a strong coupling regime, it should be possible to take a sensible decoupling limit as follows: 
$L \rightarrow \infty,~m_{\rm Pl} \rightarrow \, \infty$, keeping $\Lambda_{\rm strong}\, =\,$fixed.  The only non-linearities that
survive this limit are those associated with the helicity-0 component. Formally generalizing the decoupling prescription of~\cite{strong}
to the case of generic $\alpha$, we obtain the following equations for the helicity-2 and helicity-0 components
 \be
\label{hdecres}
\lf\mathcal{E}\tilde{h}\rt_{\mu\nu} \, =  \, -{T_{\mu\nu}  \over m_{\rm Pl}}\,,
\ee
and 
\be
\label{chidecres}
\square \chi \, 
 - {1 \over \Lambda_{\rm strong}^{5\, -\, 4\alpha}}  \left ( 3 \square \left(\square^{1-\alpha} \chi\right)^2  \, - \, \square \left({\partial_\mu\partial_\nu \over \square^{\alpha}}\chi\right)^2\, - \, 2 \partial^{\mu}\partial^{\nu} \left({\partial_{\mu}\partial_{\nu} \over \square^{\alpha}} \chi \square^{1-\alpha} \chi\right) \right) \, =  \, -{T \over m_{\rm Pl}} \,,
\ee
 where we have once assumed canonical normalization for the metric fields.
 
 Focusing on pure vacuum energy, $T_{\mu\nu} = -\Lambda\eta_{\mu\nu}$, we now wish to determine which values of $\alpha$ can degravitate the cosmological constant. 
 A necessary and sufficient condition for degravitation is that up to a trivial gauge choice in $\tilde{h}_{\mu\nu}$, 
 the solution~(\ref{hdecres}) and~(\ref{chidecres}) must be given by ~(\ref{hcc1}) and~(\ref{chicc1}), respectively:
\bea
\tilde{h}_{\mu\nu} \,& = & \, -{\Lambda \over 12 m_{\rm Pl}} \eta_{\mu\nu} x_{\alpha}x^{\alpha}\,;\\
\chi \, & = & \, {\Lambda \over 2m_{\rm Pl}}  x_{\alpha}x^{\alpha}\,.
\label{flatsoln}
\eea
For $\tilde{h}_{\mu\nu}$ this is automatically the case, since~(\ref{hdecres}) is the same as before --- see~(\ref{hdecmass}). 
For $\chi$,  however, the above condition implies that the non-linear term in (\ref{chidecres}) must identically vanish 
for~(\ref{flatsoln}).  This will be the case provided that
\begin{equation}
\label{conx}
{ \partial_{\mu}\partial_{\nu}\partial_{\beta}  \over \square^{\alpha}} \, x^2 \, = \, 0, 
\end{equation}
which in turn demands that 
\begin{equation}
\alpha \, < \, \frac{1}{2} \;.  
\label{alpharange}
\end{equation}
Thus, not every theory in the allowed range $0\,  < \, \alpha  \, < \, 1$ can degravitate the cosmological constant. The bound~(\ref{alpharange}) includes massive gravity, consistent with our previous analysis. Fortunately, the bounds $\alpha < 1/2$ and $0<\alpha < 1$ based respectively on degravitation and unitarity are not mutually exclusive.

A key question, however,  is whether the decoupling limit captures the essence of the degravitation 
phenomenon.  If it does, then there are good news and bad news. The good news is that 
degravitation of the cosmological term becomes a must, at least for a large class of theories. 
The bad news is that degravitation may go hand in hand with instabilities. It could be however that the instability is
an artefact of the decoupling limit and is absent at finite $L$. Finally, the question remains whether sensible theories within the interval  $0 \, <\, \alpha  < 1/2$ do exist from a
fundamental point of view.

\section{Filter properties in Cosmological context}

Let us go back to finite filtering scale and study degravitation in a cosmological framework. In Sec.~\ref{degravfilter} we investigated the range of validity of~(\ref{central}) as an effective filtering equation, in the context of abelian non-linear gravity.
Here we assume that similar lessons apply to fully covariant theories and proceed to study degravitation at the non-linear level using~(\ref{central}),
first in massive gravity~\footnote{A similar analysis was done together with Michele Redi in an independent unpublished study. We thank him for valuable discussions.} then using a DGP-like filter. To recap the picture,~(\ref{central}) is to be understood as the non-linear completion of the linearized equation for the helicity-2 part
$\tilde{h}$, after having integrated out the extra polarization states. The latter are strongly coupled as long as the universe is within in its own $r_\star$, corresponding to $HL\gg 1$, where $L^{-1}$ is the graviton mass or the peak of the resonance. Thus the filtering described below can only be trusted as long as $HL \gg 1$. After $H$ drops to $\sim L$, the extra states must be taken into the account. The most interesting possibility is if the latter stop the degravitation process and leave a small effective cosmological term of order $L^{-2}$.

\subsection{Massive Gravity-Type Filter}

For massive gravity-type filter, the trace of~(\ref{central}) for a vacuum energy source is given by
\be
\left(1-\frac{m^2}{\Box}\right)R(t) = 2\Lambda\,,
\label{Rm}
\ee
where $\Box = \partial_t^2 + 3H\partial_t$ in FRW coordinates. To study degravitation of vacuum energy dynamically, we seek solutions
which at early times, $t\ll m^{-1}$, are approximately de Sitter, while at late times, $t\gg m^{-1}$, asymptote to flat space. Thus the derivation below closely
parallels the linear and non-linear solutions in massive gravity obtained in Secs.~\ref{spin2} and~\ref {degravfilter}, respectively.

As mentioned above, neglecting the contribution from extra states is a good approximation if $m \ll H$, which means degravitation is slow on a Hubble time.
Thus $H$ evolves adiabatically and can be treated as constant in solving~(\ref{Rm}). The solution for $R$ is then
\be
R = - \int dt' K(t-t')T(t') = 2\Lambda \int dt'K(t-t') \,,
\ee
with kernel given by 
\be
K(t) = \delta(t) + \frac{m^2}{2\pi}\int d\om \frac{e^{i\om t}} {\om^2 - 3Hi\om -m^2}\,.
\label{Km}
\ee
This integrand has two poles, both in the upper $\om$-plane, and thus the filter is causal with appropriate choice of contour.
The end result for $R(t>0)$ is a sum of growing and decaying exponentials, which greatly simplifies in the adiabatic limit $H/m\ll 1$:
\be
\frac{R(t)}{2\Lambda} \approx \exp\left(-\frac{m^2t}{H}\right)\,.
\label{Rmsoln}
\ee
As advocated, this reduces to de Sitter at early times. At late times, however, the backreaction of the source become exponentially small, and the geometry asymptotes to flat space. Restoring the time-dependence of $H$, and using $R(t)\approx 12H^2(t)$, one can then solve~(\ref{Rmsoln}) for the evolution of the Hubble parameter.

\subsection{DGP-Type Filter}
\label{dgptype}

Next we consider a DGP-like filter: $m^2(\Box) = L^{-1}\sqrt{-\Box}$, corresponding to $\alpha =1/2$. Although we have argued in the previous section that theories with $\alpha =1/2$ cannot
degravitate vacuum energy in the strict decoupling limit, it is not clear what happens for finite filter scale. Moreover, even though degravitation does not take place in full DGP, the latter may not be the unique theory with $\alpha = 1/2$ in the linearized theory. Our point here is simply to show that a DGP-like filter can degravitate the cosmological constant.

The filter operator can be defined through the spectral representation for the DGP propagator:
\be
\frac{1}{-\Box + L^{-1}\sqrt{-\Box}} =  \frac{2L}{\pi } \int_0^\infty \frac{dm}{1+ m^2L^2}\frac{1}{-\Box+m^2}\,,
\ee
with the spectral density $\rho(m) = L/(1+m^2L^2)$. This expression holds for any background, including FRW, as shown in detail in Appendix~\ref{ap1}. The solution for $R$ is then just a superposition of the massive solutions, appropriately weighted by the spectral density:
\be
\frac{R(t)}{2\Lambda} = \frac{2L}{\pi } \int_0^\infty \frac{dm}{1+ m^2L^2}R^{(m)}\,,
\ee
where $R^{(m)}$ solves~(\ref{Rm}). 

Since the integral runs over modes of arbitrarily large mass, the approximation $H/m\gg 1$ used in~(\ref{Rmsoln}) is bound to break down at the upper end of the spectrum. However, the DGP spectral density has an exponentially small tail in this limit, and thus it is a good approximation to substitute~(\ref{Rmsoln}) for all modes:
\be
\frac{R(t)}{2\Lambda} \approx  \frac{2\pi}{L} \int_0^\infty \frac{dm}{1+ m^2L^2}\exp\left(-\frac{m^2t}{H}\right) = e^{t/HL^2}{\rm Erfc}\left(\sqrt{\frac{t}{HL^2}}\right)\,.
\ee
See Appendix~\ref{ap2} for the exact derivation of the solution, which does not rely on the $H/m\gg 1$ approximation. At early times, $t \ll L$, we have
\be
\frac{R(t)}{2\Lambda} \approx  1-\frac{2}{\sqrt{\pi}}\sqrt{\frac{t}{HL^2}}\,,
\ee
which is just de Sitter plus a small correction. Moreover, using the asymptotic form of the complementary error function, ${\rm Erfc}(x)\sim e^{-x^2}/x$ for large $x$, we see that vacuum energy asymptotically degravitates as
\be
\frac{R(t)}{2\Lambda} \rightarrow  \frac{1}{\sqrt{\pi}}\sqrt{\frac{HL^2}{t}} \sim t^{-1/2}\,.
\label{late}
\ee

\subsection{General Filter}

It is easy to generalize the above derivativation to arbitrary filters. As long as the corresponding spectral density is peaked around some mass scale $L^{-1}$ such that $HL\gg 1$, with exponentially small tail for $m\gsim H$, then the solution can be approximated by
\be
\frac{R(t)}{2\Lambda} \approx  \int_0^\infty dm\;L \frac{\rho(m)}{L}\exp\left(-\frac{m^2t}{H}\right) \,.
\ee
At late times for instance, the integrand is significant only for $mL\ll 1$, and thus we can approximate $\rho(m) \approx \rho(0)$. This gives
\be
\frac{R(t)}{2\Lambda} \rightarrow \frac{\rho(0)}{L}\frac{\sqrt{\pi}}{2}\sqrt{\frac{HL^2}{t}}\,.
\ee
For DGP, $\rho(0) = 2L/\pi$, and we recover~(\ref{late}). Thus the $t^{-1/2}$ fall-off at late times is generic for a broad class of filters.

\acknowledgments

We thank Gregory Gabadadze and Michele Redi for valuable discussions. 
This work of G.D. is supported in part by David and Lucile Packard Foundation Fellowship for Sience and Engineering, and by NSF grant PHY-0245068. The research of S.H. and J.K. at Perimeter Institute is supported in part by the Government of Canada through NSERC and by the Province of Ontario through MRI. G.D. would like to thank Perimeter Institute where the part of this work was done. S.H. and J.K. are grateful to the Center for Cosmology and Particle Physics at NYU for their hospitality.

\begin{appendix}
     
\section{defining $\sqrt{\square}$} \label{ap1}
   
In this section we provide a rigorous definition of the operator $\sqrt{\hat{\mathcal{O}}}$, for generic ordinary
differential operator $\hat{\mathcal{O}}$ depending on $t$ and $\partial_t$. For instance, in the
derivation of Sec.~\ref{dgptype}, we have  $\hat{\mathcal{O}} = \partial_t^2 + 3H(t)\partial_t$.
  
In the case where $\hat{\mathcal{O}}$ is independent of $t$ and is therefore a function of derivatives only,  the square root operator can be defined in the usual way through
Fourier transform. For example, its action on a function $f(t)$ can be defined in the following way
\begin{equation}
\label{definition}
\sqrt{\hat{\mathcal{O}}(\partial_t)} \, f(t) \, = \, \int \, {d\omega \over 2\pi}  {\rm e}^{i \omega} \mathcal{O}(i\omega) \, \tilde{f}(\omega)\,.
\end{equation} 
In particular,  we could apply  this definition to $\sqrt{\partial_t^2 +3H\partial_t}$ for  $H =$ constant. 

Here we wish to generalize the notion of $\sqrt{\hat{\mathcal{O}}}$ to the case of time-dependent
operator. To do so, let us introduce an auxiliary space coordinate  $y$, and consider the following
equation 
\begin{equation}
\label{yequation}
\left (\hat{\mathcal{O}} \, - \, \partial_y^2 \, + \, L \, \delta(y) \hat{\mathcal{O}} \right) 
Q(t,y)\, = \, \delta(y)\delta(t) \,.
\end{equation} 
The solution of this equation can be taken to be the form
\begin{equation}
\label{qexpansion}
Q(y,t) \, = \, \int\, dm \,  \psi^{(m)}(y) \, Q^{(m)}(t) \,,
\end{equation}
where the functions $\psi^{(m)}(y)$ satisfy
\begin{equation}
\label{yequation2}
\left (m^2\, + \, \partial_y^2 \, + \, L\,  \delta(y)\, m^2 \right) 
\psi^{(m)}(y)\, = \, 0\,.
\end{equation} 
The $\psi^{(m)}$'s form a complete and orthonormal set, with the norm given by
\begin{equation}
\label{norm}
\int dy \, \psi^{(m)}(y) (1 \, + \, L \, \delta(y)) \psi^{(m)}(y) \, =\, \delta(m\, - \, m')\,.
\end{equation}

Substituting the expansion~(\ref{qexpansion}) into~(\ref{yequation}), 
multiplying both sides by $\psi^{(m')}$, and integrating over $y$, we find that the functions $Q^{(m)}(t)$ satisfy
\begin{equation}
\label{yequation3}
\left (\hat{\mathcal{O}} \, + \, m^2 \right) 
Q(t,y)\, = \, \psi^{(m)}(0)\delta(t) \,.
\end{equation} 
Hence the solution of (\ref{yequation}) is given by 
\begin{equation}
\label{qexpansion2}
Q(y,t) \, = \, \int\, dm \,  \psi^{(m)}(y)\psi^{(m)}(0) {1 \over \hat{\mathcal{O}} \, + \, m^2}
\delta(t) \,.
\end{equation}

Now a key realization is that  $Q(t) \equiv Q(t,y=0)$ is the Green's function 
of the $\hat{\mathcal{O}} + \sqrt{\hat{\mathcal{O}}}/L$ operator. Using the fact that $\psi^{(m)}(0)\psi^{(m)}(0) \, = \, ((mL)^2 \, + \, 1)^{-1} $, we get 
\begin{equation}
\label{qexpansion3}
Q(t) \, = \, \int\, {dm \over (mL)^2 \, + \, 1}  {1 \over \hat{\mathcal{O}} \, + \, m^2}
\delta(t) \,,
\end{equation}
which in operator language translates to
\begin{equation}
\label{sqrtdef}
{1 \over L \, \hat{\mathcal{O}} \, + \,  \sqrt{\hat{\mathcal{O}}}} \, 
 = \, \int\, {dm \over (mL)^2 \, + \, 1}  {1 \over \hat{\mathcal{O}} \, + \, m^2} \,.
\end{equation}

Note that in case when  $\hat{\mathcal{O}}$ only depends on derivatives,  
this definition coincides with the conventional one using Fourier transform, as can be seen using the general
relation
\begin{equation}
\label{qeq}
 {1\over  Lq^2  \, + \, \sqrt{q^2}} \, = \, \int  {dm \over (mL)^2 \, + \, 1} {1 \over q^2 \, + \, m^2}\,.
\end{equation}

\section{Exact Solution for DGP-like Filter} \label{ap2}

Here we provide an exact solution for degravitation with DGP-like filter, generalizing the solution of Sec.~\ref{dgptype}. The trace of the filtering equation in this case is given by
\be
\left(1 + \frac{1}{L\sqrt{-\Box}}\right)R(t) = 2\Lambda\,.
\ee
Assuming $H$ is approximately constant, the solution for $R$ can be written as 
\be
R = -2\Lambda \int dt' K(t-t')\,,
\label{R}
\ee
where the kernel $K$ is given as usual in terms of a Fourier integral
\be
K(t) = \delta(t) - \frac{1}{2\pi i L }\int d\om e^{i\om t} \frac{\sqrt{\om^2 - 3Hi\om} +L^{-1}i}{(\om-\om_+)(\om -\om_-)}  \equiv \delta(t) + \cI(t)\,,
\label{KDGP}
\ee
where
\be
\om_\pm = \frac{3Hi}{2}\left(1\pm\sqrt{1+\frac{4}{9H^2L^2}}\right)\,.
\ee
Equation~(\ref{KDGP}) is the DGP-like analogue of~(\ref{Km}) for massive gravity.

\begin{figure}[ht]
\centering
\includegraphics[width=100mm]{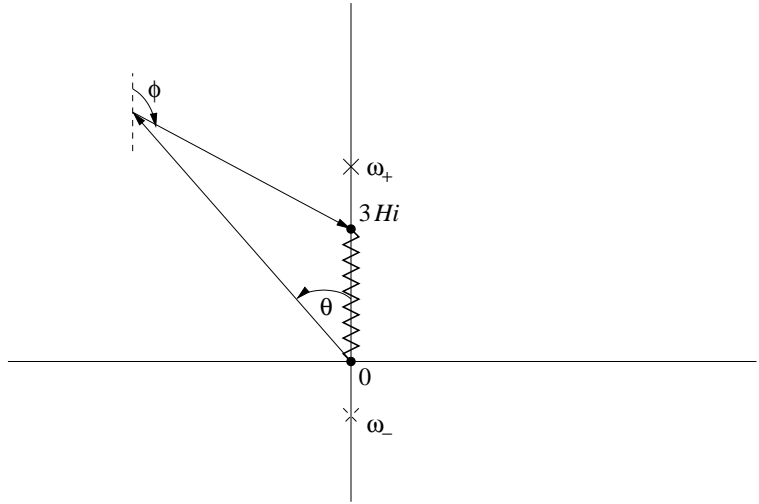}
\caption{Conventions for the branch cut in the integrand for $\cI$. This function has two poles, at $\om_+$ and $\om_-$, which lie in the upper and lower half plane, respectively. The pole in the lower half plane, however, lies on the second Riemann sheet and can be accessed by going under the branch cut.}
\label{complexplane}
\end{figure}

It remains to calculate $\cI$. Its integrand has branch points at $\om = 0$ and $\om = 3Hi$, so it is natural to choose the branch cut to extend between these points, as shown in Fig.~\ref{complexplane}. The representation of the square root is taken to be 
\be
\sqrt{\om^2 - 3Hi\om}  = - \sqrt{|\om|}e^{i\theta/2} \sqrt{|\om-3Hi|} e^{i\phi/2}\,,
\label{sq}
\ee
where the angles $\theta$ and $\phi$ are defined in the Figure. 

\begin{figure}[ht]
\centering
\includegraphics[width=100mm]{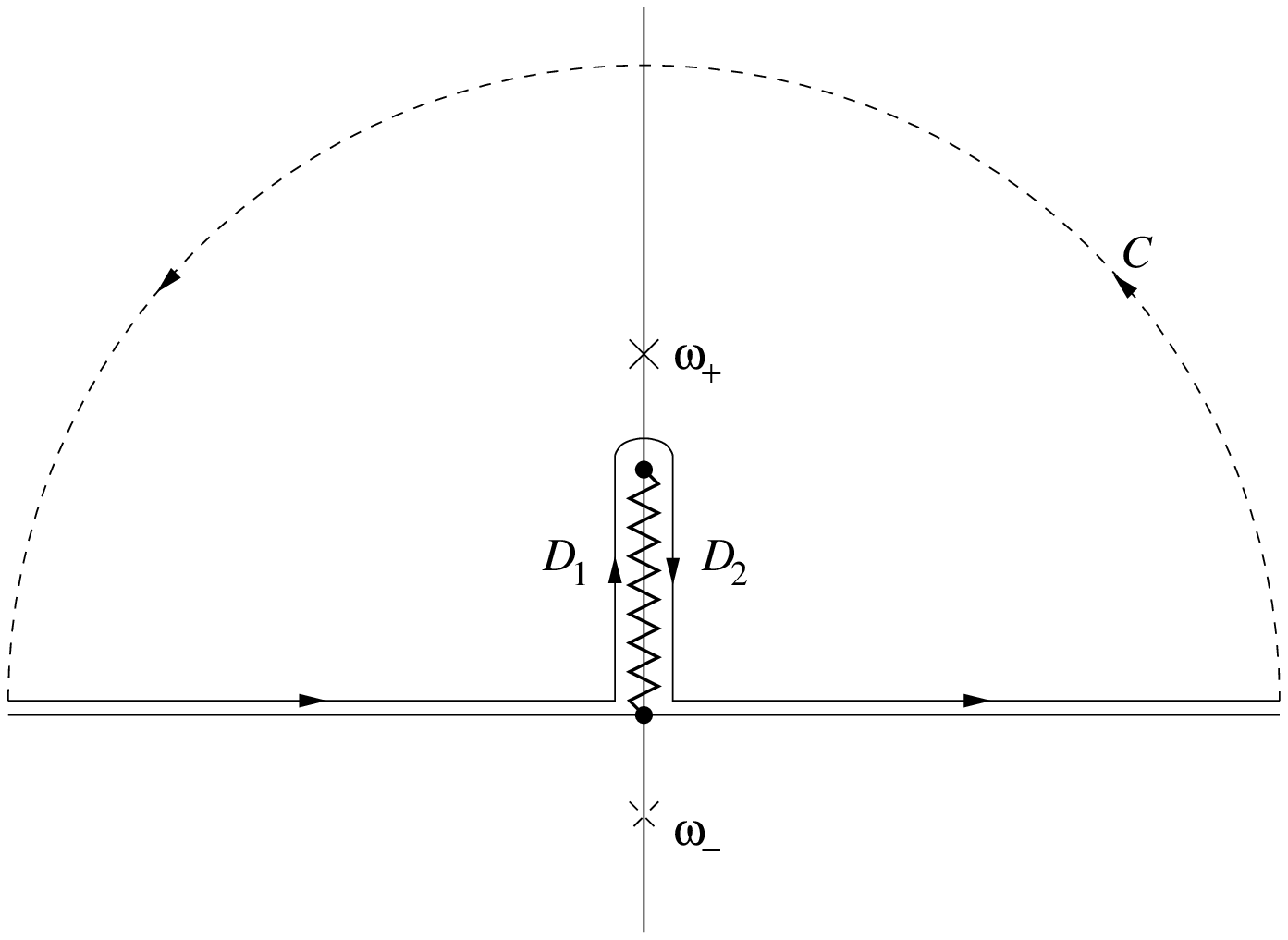}
\caption{Choice of contour for $\cI(t>0)$.}
\label{contour}
\end{figure}

Furthermore, the integrand has poles at $\om=\om_\pm$. Note, however, that the pole at $\om_-$ actually lies on the second Riemann sheet. 
For $t>0$ we choose the contour $\cC$ shown in Fig.~\ref{contour} and obtain
\be
\cI = -\frac{2}{3HL^2\sqrt{1+\frac{4}{9H^2L^2}}}\exp\left[-\frac{3Ht}{2}\left(1+\sqrt{1+\frac{4}{9H^2L^2}}\right)\right] + \cQ\,,
\ee
where
\be
\cQ \equiv \frac{1}{2\pi i L }\sum_{i=1}^2 \int_{\cD_i}dz \frac{e^{iz t}\sqrt{z^2 - 3Hiz}}{(z-z_+)(z-z_-)} = \frac{1}{\pi i L } \int_0^{3Hi} dz \frac{e^{iz t}\sqrt{z^2 - 3Hiz}}{(z-z_+)(z-z_-)}\,.
\ee
In the last step we have used~(\ref{sq}) to deduce that $\sqrt{z^2 - 3Hiz} = - \sqrt{|z||z-3Hi|}$ along $\cD_1$ and $ + \sqrt{|z||z-3Hi|}$ along $\cD_2$. Changing variables to $ x\equiv z/3Hi$, this reduces to
\be
\cQ = \frac{1}{\pi L}\int_0^1 dx\frac{e^{-3Htx}\sqrt{x}\sqrt{1-x}}{x^2-x-(3HL)^{-2}}\,,
\ee
which can be evaluated numerically. Hence our final expression for the kernel is
\bea
\nonumber
K(t) &=& \delta(t)  - \frac{2}{3HL^2\sqrt{1+\frac{4}{9H^2L^2}}}\exp\left[-\frac{3Ht}{2}\left(1+\sqrt{1+\frac{4}{9H^2L^2}}\right)\right]  \\
&+&  \frac{1}{\pi L}\int_0^1 dx\frac{e^{-3Htx}\sqrt{x}\sqrt{1-x}}{x^2-x-(3HL)^{-2}}\,.
\label{Kfinal}
\eea

\end{appendix}

\pagebreak

\end{document}